\documentclass[aps,floatfix,lengthcheck,showpacs,amssymb,amsmath,amsfonts,
twocolumn,nofootinbib,prd]{revtex4-1}
\usepackage[utf8]{inputenc}
\usepackage{tabularx}
\usepackage{color}
\usepackage[table]{xcolor}
\usepackage{graphicx}
\usepackage{float}
\usepackage{subfigure}
\usepackage{acronym}
\usepackage{tikz}
\usetikzlibrary{shapes.geometric, arrows, calc, fit, positioning, chains, arrows.meta}
\usepackage{hyperref}
\hypersetup{
    colorlinks=true,
    linkcolor=blue,
    filecolor=magenta,      
    urlcolor=cyan,
}
\usepackage{verbatim}
\usepackage[binary-units=true]{siunitx}
\usepackage{bm}
\usepackage{mathtools}
\usepackage{rotating}
\usepackage[toc, nonumberlist, acronym]{glossaries}
\glsdisablehyper
\usepackage{multirow}
\usepackage{xspace}

\graphicspath{images/}

\DeclareSIUnit\parsec{pc}

\definecolor{Gray}{gray}{0.9}

\begin{document}

\newacronym{mlgwsc1}{MLGWSC-1}{first Machine Learning Gravitational-Wave Mock Data Challenge}
\newacronym{bbh}{BBH}{Binary Black Hole}
\newacronym{bns}{BNS}{Binary Neutron Star}
\newacronym{nsbh}{NSBH}{Neutron Star Black Hole}
\newacronym{gw}{GW}{Gravitational Wave}
\newacronym{o1}{O1}{first observing run}
\newacronym{o2}{O2}{second observing run}
\newacronym{o3}{O3}{third observing run}
\newacronym{o3a}{O3a}{first half of the third observing run}
\newacronym{o3b}{O3b}{second half of the third observing run}
\newacronym{o4}{O4}{fourth observing run}
\newacronym{psd}{PSD}{Power Spectral Density}
\newacronym{roc}{ROC}{Receiver Operating Characteristic}
\newacronym{far}{FAR}{False-Alarm Rate}
\newacronym{lvk}{LVK}{LIGO-Virgo-KAGRA}
\newacronym{gwosc}{GWOSC}{Gravitational Wave Open Science Center}
\newacronym{snr}{SNR}{Signal to Noise Ratio}
\newacronym{usr}{USR}{Unbounded Softmax Replacement}

\title[]{
MLGWSC-1: The first Machine Learning Gravitational-Wave Search Mock Data Challenge
}

\newcommand{\jena}{TPI FSU Jena\xspace}
\newcommand{\virgo}{Virgo-AUTh\xspace}
\newcommand{\cnn}{CNN-Coinc\xspace}
\newcommand{\pycbc}{PyCBC\xspace}
\newcommand{\cwb}{cWB\xspace}
\newcommand{\mfcnn}{MFCNN\xspace}

\author{
    Marlin B. Sch{\"a}fer$^{1,2}$, 
    Ond{\v{r}}ej Zelenka$^{3,4}$, 
    Alexander H. Nitz$^{1,2}$, 
    He Wang$^{5}$, 
    Shichao Wu$^{1,2}$, 
    Zong-Kuan Guo$^{5}$, 
    Zhoujian Cao$^{6}$, 
    Zhixiang Ren$^{7}$, 
    Paraskevi Nousi$^{8}$, %
    Nikolaos Stergioulas$^{9}$, 
    Panagiotis Iosif$^{10, 9}$, 
    Alexandra E. Koloniari$^{9}$, %
    Anastasios Tefas$^{8}$, %
    Nikolaos Passalis$^{8}$, %
    Francesco Salemi$^{11, 12}$, 
    Gabriele Vedovato$^{13}$, 
    Sergey Klimenko$^{14}$, %
    Tanmaya Mishra$^{14}$, 
    Bernd Br{\"u}gmann$^{3,4}$, 
    Elena Cuoco$^{15, 16, 17}$, 
    E. A. Huerta$^{18, 19}$, 
    Chris Messenger$^{20}$, 
    Frank Ohme$^{1,2}$
}
\address{$^1$Max-Planck-Institut f{\"u}r Gravitationsphysik,
         Albert-Einstein-Institut, D-30167 Hannover, Germany}
\address{$^2$Leibniz Universit{\"a}t Hannover, D-30167 Hannover, Germany}
\address{$^3$Friedrich-Schiller-Universit{\"a}t Jena, D-07743 Jena, Germany}
\address{$^4$Michael Stifel Center Jena, D-07743 Jena, Germany}
\address{$^5$CAS Key Laboratory of Theoretical Physics, Institute of Theoretical Physics, Chinese Academy of Sciences, Beijing 100190, China}
\address{$^6$Department of Astronomy, Beijing Normal University, Beijing 100875, China}
\address{$^7$Peng Cheng Laboratory, Shenzhen, 518055, China}
\address{$^8$Department of Informatics, Aristotle University of Thessaloniki, GR-54124 Thesssaloniki, Greece}
\address{$^9$Department of Physics, Aristotle University of Thessaloniki, GR-54124 Thessaloniki, Greece}
\address{$^{10}$GSI Helmholtz Center for Heavy Ion Research, Planckstra{\ss}e 1, 64291 Darmstadt, Germany}
\address{$^{11}$Universit\`a di Trento, Dipartimento di Fisica, I-38123 Povo, Trento, Italy}
\address{$^{12}$INFN, Trento Institute for Fundamental Physics and Applications, I-38123 Povo, Trento, Italy}
\address{$^{13}$INFN, Sezione di Padova, I-35131 Padova, Italy}
\address{$^{14}$Department of Physics, University of Florida, PO Box 118440, Gainesville, FL 32611-8440, USA}
\address{$^{15}$European Gravitational Observatory (EGO), I-56021 Cascina, Pisa, Italy}
\address{$^{16}$Scuola Normale Superiore, Piazza dei Cavalieri 7, I-56126 Pisa, Italy}
\address{$^{17}$INFN, Sezione di Pisa, Largo Bruno Pontecorvo, 3, I-56127 Pisa, Italy}
\address{$^{18}$Data Science and Learning Division, Argonne National Laboratory, Lemont, Illinois 60439, USA}
\address{$^{19}$Department of Computer Science, University of Chicago, Chicago, Illinois 60637, USA}
\address{$^{20}$SUPA, School of Physics and Astronomy, University of Glasgow, Glasgow G12 8QQ, United Kingdom}

\begin{abstract}
We present the results of the first Machine Learning Gravitational-Wave Search Mock Data Challenge (\acrshort{mlgwsc1}). For this challenge, participating groups had to identify gravitational-wave signals from binary black hole mergers of increasing complexity and duration embedded in progressively more realistic noise. The final of the 4 provided datasets contained real noise from the O3a observing run and signals up to a duration of 20 seconds with the inclusion of precession effects and higher order modes. We present the average sensitivity distance and runtime for the 6 entered algorithms derived from 1 month of test data unknown to the participants prior to submission. Of these, 4 are machine learning algorithms. We find that the best machine learning based algorithms are able to achieve up to $95\%$ of the sensitive distance of matched-filtering based production analyses for simulated Gaussian noise at a false-alarm rate (\acrshort{far}) of one per month. In contrast, for real noise, the leading machine learning search achieved $70\%$. For higher \acrshort{far}s the differences in sensitive distance shrink to the point where select machine learning submissions outperform traditional search algorithms at \acrshort{far}s $\geq 200$ per month on some datasets. Our results show that current machine learning search algorithms may already be sensitive enough in limited parameter regions to be useful for some production settings. To improve the state-of-the-art, machine learning algorithms need to reduce the false-alarm rates at which they are capable of detecting signals and extend their validity to regions of parameter space where modeled searches are computationally expensive to run. Based on our findings we compile a list of research areas that we believe are the most important to elevate machine learning searches to an invaluable tool in gravitational-wave signal detection.
\end{abstract}

\maketitle

\section{Introduction}
The first gravitational-wave (\acrshort{gw}) observation on September 14, 2015~\citep{LIGOScientific:2016aoc} achieved by the LIGO and Virgo Collaboration~\citep{LIGOScientific:2014pky, TheVirgo:2014hva} started the era of \acrshort{gw} astronomy. During the first observing run (\acrshort{o1}) two more \acrshort{gw}s from coalescing binary black holes (\acrshort{bbh}s) were detected. The second observing run (\acrshort{o2}) saw $\mathcal{O}(10)$ additional confident \acrshort{bbh} detections as well as the first detection of a binary neutron star (\acrshort{bns}) merger~\citep{LIGOScientific:2017vwq, LIGOScientific:2018mvr, Nitz:2018imz, Nitz:2020oeq, Venumadhav:2019tad, Venumadhav:2019lyq}. The third observing run (\acrshort{o3}) was split into two parts, \acrshort{o3a} and \acrshort{o3b}. During \acrshort{o3a} a further $\mathcal{O}(40)$ \acrshort{bbh}s as well as a second \acrshort{bns} merger were reported~\citep{LIGOScientific:2020ibl, LIGOScientific:2021usb, Nitz:2021uxj}. \acrshort{o3b} added another $\mathcal{O}(40)$ \acrshort{bbh} events as well as finding the first two confident detections where the component masses are consistent with the merger of a neutron star black hole system (\acrshort{nsbh})~\citep{LIGOScientific:2021djp, Nitz:2021zwj}. The fourth observing run (\acrshort{o4}) is scheduled to begin in spring of 2023 and is expected to significantly increase the volume from which sources can be detected~\citep{Abbott:2020qfu, cahillane2022review}.

\acrshort{gw} signals are commonly identified in the background noise of the detectors using matched filtering~\citep{LIGOScientific:2021djp, Messick:2016aqy, DalCanton:2020vpm, Adams:2015ulm}. Matched filtering compares pre-computed models of expected signals, known as templates, with the data from the detectors~\citep{Allen:2005fk}. When a model matches the data to a pre-defined degree and data-quality requirements are met, a candidate detection is reported.
Loosely modelled searches~\citep{Klimenko:2005xv, Klimenko:2015ypf}, which look for coherent excess power in multiple detectors, are also employed by the LIGO-Virgo-KAGRA collaboration (\acrshort{lvk}) to find potential signals.

The rate of detections has drastically increased from \acrshort{o1} to \acrshort{o3}. This increase was enabled by continued detector upgrades at the two advanced LIGO observatories in Hanford and Livingston~\citep{LIGOScientific:2014pky}, as well as sensitivity improvements for the advanced Virgo detector~\citep{TheVirgo:2014hva}. With the entry into service of Kagra~\citep{KAGRA:2018plz} a fourth observatory joined the network of ground based \acrshort{gw}-detectors towards the end of \acrshort{o3}. The rate of detections is expected to further increase during \acrshort{o4} as the sensitivity of the detectors improves and the volume from which sources can be detected grows.

With an increasing rate of detections, it is likely that systems with unexpected physical properties will be observed more frequently in the future. Optimally searching for these is a challenge for matched filtering based searches, where the computational cost scales linearly with the number of templates used. The inclusion of effects such as precession, eccentricity, or higher order modes requires millions of templates to not miss potential signals~\citep{Harry:2016ijz, Harry:2017weg, Dhurkunde:2022aek} and thus are computationally prohibitive, especially when real-time alerts should be issued. Loosely modeled searches are inherently capable of detecting arbitrary sources at a fixed computational cost but are prone to miss more signals due to their lower sensitivity in parameter regions where accurate models exist.

In recent years, machine learning has been applied in many scientific fields to enable or improve research into computationally expensive topics~\citep{Deiana:2021niw}. Some examples include the prediction of protein structure used in pharmaceutical studies~\citep{Alquraishi:2021aaa}, improvements to material composition and synthesis~\citep{Butler:2018aaa}, or event reconstruction at the Large Hadron Collider~\citep{Gray:2020mcm}. There is also ongoing research into using neural networks to discover closed form expressions from raw data~\citep{Cranmer:2020wew} or optimizing machine learning algorithms to take advantage of physical symmetries of the underlying problem~\citep{Smidt:2020tuy, Oxley:2021aaa, Dax:2021myb}.

More relevant to this work, machine learning algorithms have also started to be explored as alternative algorithms for many \acrshort{gw} data-analysis tasks. These include detector glitch classification~\citep{Zevin:2016qwy, George:2017fbn, Bahaadini:2018git}, parameter estimation~\citep{Chua:2019wwt, Gabbard:2019rde, Chatterjee:2019avs, Dax:2021tsq, McLeod:2022ccr}, continuous \acrshort{gw} detection~\citep{Dreissigacker:2019edy, Morawski:2019awi, Miller:2019jtp, Dreissigacker:2020xfr, Beheshtipour:2020zhb, Beheshtipour:2020nko, Yamamoto:2022adl}, enhancements for existing pipelines~\citep{Chatterjee:2019gqr, Cabero:2020eik, Jadhav:2020oyt, Cabero:2020eik, Williams:2021qyt, Mishra:2021tmu, Lopez:2021ikt, Mishra:2022ott, McIsaac:2022odb}, surrogate waveform models~\citep{Khan:2020fso, Nousi:2021arn, Fragkouli:2022lpt}, as well as various signal detection algorithms~\citep{George:2016hay, George:2017pmj, Gabbard:2017lja, Gebhard:2019ldz, Krastev:2019koe, Krastev:2020skk, Wei:2020ztw, Schafer:2020kor, Wei:2020sfz, Wei:2020xrl, Huerta:2020xyq, Morawski:2021kxv, Schafer:2021fea, Schafer:2021cml, Khan:2021vbf, Ruan:2021fxq, Chaturvedi:2022suc, Baltus:2022pep, Moreno:2021fvp, Bacon:2022lsm}. For a summary of many methods we refer the reader to \citep{Cuoco:2020ogp, Huerta:2021ybd}. In this work we focus solely on detection algorithms for \acrshort{bbh} \acrshort{gw} signals, which have been the most commonly observed type of sources to date~\citep{LIGOScientific:2020ibl, LIGOScientific:2021usb, Nitz:2021uxj}. These signals are the easiest to detect for machine learning algorithms due to their short duration.

Many of the works considering the usage of machine learning for \acrshort{gw} signal detection are difficult to cross-compare. Most algorithms target different datasets and derived metrics are often motivated more by machine learning practices than by state-of-the-art \acrshort{gw} searches. It is, therefore, hard to pinpoint exactly how capable machine learning search algorithms currently are and where the main difficulties arise. To achieve the goal of an objective characterization of machine learning \acrshort{gw} search capabilities, a common ground for comparison is required.

Here we present the results of the first Machine Learning Gravitational-Wave Search Mock Data Challenge (\acrshort{mlgwsc1}). In an attempt to provide a common ground of comparison for different algorithms and in preparation of \acrshort{o4}, we have calculated sensitive distances from 6 different submissions calculated on datasets of one month duration to collect and compare a suite of searches. We want to motivate the utilization of machine learning based searches in a production setting by providing a definitive resource to allow for easy comparison between different algorithms, be it machine learning based, matched filtering based, or completely unmodeled. This challenge is the first of its kind\footnote{There has previously been a public Kaggle challenge \citep{kaggle_challenge}. First in the sense of this paper refers to our setup of providing continuous data.} and hopefully more will be held in the future, expanding to more difficult scenarios.

The mock data used in this challenge consists of 4 datasets containing noise of increasing realism and signals with increasing complexity for the two detectors LIGO Hanford and LIGO Livingston~\citep{LIGOScientific:2014pky}. The final dataset challenges participants to identify \acrshort{gw}s from spinning \acrshort{bbh}s with a duration of up to \SI{20}{\second} added to real detector noise from \acrshort{o3a}. The signals also take precession effects and higher order modes into account.

Submissions are evaluated on mock data of one month duration for each of the four datasets. We calculate sensitive distances for each algorithm and estimate the computational efficiency based on the runtime. The final dataset should provide an accurate picture of the possible real-world performance these algorithms can achieve. However, we note that direct comparison of the runtime performance of the different algorithms is complicated by differing hardware usage and optimization.

We find that machine learning algorithms are already competitive with state-of-the-art searches on simulated data containing injections drawn from the limited parameter space covered by this challenge. The most sensitive machine learning algorithm manages to retain $\geq 93\%$ of the sensitive distance measured for the \pycbc pipeline~\citep{Nitz:2021zwj} on Gaussian background data down to a false-alarm rate (\acrshort{far}) of $1$ per month. For higher \acrshort{far}s the separation between the approaches generally shrinks.

Most machine learning searches, as tested here, are less sensitive on real noise than on simulated data. The traditional algorithms handle this transition better. As a consequence, the most sensitive machine learning algorithm retains $\geq 70\%$ of the sensitive distance of the \pycbc search down to a \acrshort{far} of 1 per month. However, the sensitivity achieved of machine learning algorithms on real data is still substantial and shows that they are capable of rejecting non-Gaussian noise artifacts without any hand-tuned glitch classification.

From the evaluation of the different datasets we conclude that the main difficulties for current machine learning algorithms are the ability to analyze the consistency of detected signals between detectors and the maximum duration of signals that can be detected. Solving these issues would allow for better performance at \acrshort{far}s $<1$ per month and enable a fast detection of potentially electromagnetic bright sources such as \acrshort{bns} or \acrshort{nsbh} mergers.

All code used in this challenge is open source and available at~\citep{github}. Therein we also collect the individual submissions by groups that have given their consent, provide the analysis results, and make available all plots used in this paper for all submissions.

This paper is structured as follows. In \autoref{sec:methods} we provide the details on the challenge, the datasets, as well as the evaluation process. All submissions are briefly introduced in \autoref{sec:submissions}. The results of the challenge and a brief discussion can be found in \autoref{sec:results}. We conclude and give an outlook into possible future work in \autoref{sec:conclusions}.

\section{Methods}\label{sec:methods}
All submissions described in \autoref{sec:submissions} are evaluated on the same datasets, and all machine learning submissions are evaluated under the same conditions. Below we describe the provided material from the challenge, the requirements for the submitted algorithms, as well as the evaluation process.

\subsection{Challenge Resources}
In this challenge participants are asked to identify \acrshort{gw} signals submerged in detector noise. To provide grounds of comparison, all submissions are evaluated on the same datasets. To allow for optimization of the submitted algorithms for the task at hand, participants had access to code that allowed them to generate arbitrary amounts of data equivalent to that used during the final evaluation of this challenge. All code used for data generation and algorithm evaluation is open source and can be found at~\citep{github}.

In particular, participants had access to the code that was used to generate the final challenge sets, but not the specific seed that was used. The specifics of the datasets are described in \autoref{sec:data}. They were also provided with the code that was used to generate the metrics we provide in this paper. Details on the metrics can be found in \autoref{sec:metrics}.

\subsection{Test Data}\label{sec:data}
The challenge provides a script to generate semi-continuous raw test data for any of the four datasets described below. It allows the user to choose a specific seed and a total duration of the output data. The code subsequently generates up to three files; the first containing pure noise, the second containing the same noise with injected \acrshort{gw} signals, and the third containing the parameters of the injected signals.

The files containing the pure noise and the noise with additive signals are of the same structure. They are HDF5 \citep{hdf5} files with two groups named ``H1'' and ``L1'' containing data from the two detectors LIGO Hanford and LIGO Livingston, respectively. Each group consists of $N$ HDF5-datasets, each holding the detector data of a single segment, as well as information on the GPS starting time of the segment, and its sampling rate. Each segment has a minimum duration of \SI{2}{\hour}, is sampled at \SI{2048}{\hertz}, and contains continuous data. The files also contain information on the meta-data used to create the file. This meta-data is removed in the final challenge sets.

We chose to split data into smaller segments of uncorrelated noise for two reasons. First, real detectors are not equally sensitive for months at a time and data quality differs to an extent where certain data cannot be used for analyses. As such, any algorithm should be able to handle gaps in the data. Second, the noise characteristic varies over time. Segmenting simulated data allows us to easily incorporate different models for the power spectrum over the duration of the data. Subsequently, the noise model can be increased in complexity for the four datasets.

Minimal pre-processing is done on the data that is handed to the submitted algorithms. We only apply a low-frequency cutoff of \SI{15}{\hertz} which is used to enable a reduction in file size for real-detector data that has to be downloaded. The low-frequency cutoff reduces the dynamic range of the data, which allows us to scale the data and cast it to lower numerical precision. Any other pre-processing is left to the algorithms and is factored into the performance evaluation. The scaling is inverted during data loading.

A larger index of the dataset signifies a greater complexity and realism of the dataset. Participants may choose to optimize for any of the 4 datasets but are only allowed to submit a single algorithm, which is subsequently tested with all 4 datasets. We do this to test the ability of the search to generalize to slightly varying conditions.

Many parameters of the injected signals are drawn from the same distributions irrespective of the dataset. A summary of these distributions can be found in \autoref{tab:shared_params}. All signals are generated using the waveform model \verb|IMRPhenomXPHM|~\citep{Pratten:2020ceb} with a lower frequency cutoff of \SI{20}{\hertz}. The waveform model was chosen for its ability to simulate both precession and higher-order modes. The merger times of two subsequent signals are seperated by a random time between \SIrange{24}{30}{\second} to avoid any overlap. We apply a taper to the start of each waveform.

\begin{figure*}
    \centering
    \includegraphics[width=0.98\textwidth]{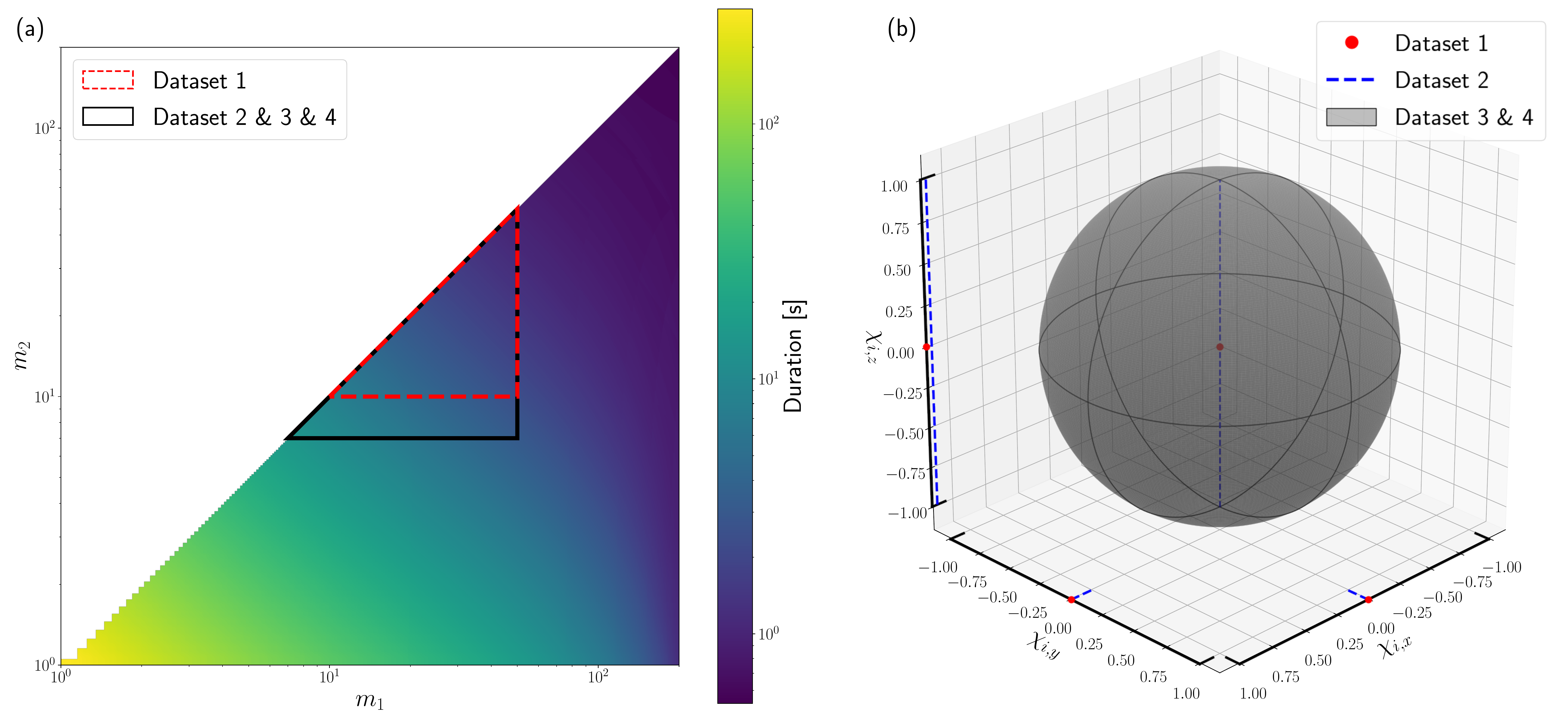}
    \caption{An illustration of the range for the intrinsic parameters covered by this challenge. The left panel (a) shows a typical range for the component masses used by state of the art searches~\citep{Nitz:2021zwj}. The color indicates the duration of the waveform from \SI{20}{\hertz}. The triangles show the parameter regions covered by this challenge. The right panel (b) shows the component-spin $\chi_i$ distribution of the different datasets in this challenge.}
    \label{fig:parameters}
\end{figure*}

In \autoref{fig:parameters} we show an overview of the intrinsic parameters used in this challenge and compare it to the parameter space searched by state of the art searches~\citep{LIGOScientific:2021djp, Nitz:2021zwj}.

\begin{table}
\centering
\begin{tabularx}{0.48\textwidth}{XX}
    \hline\hline
    Parameter & Uniform distribution \\
    \hline
    Coalescence phase & $\Phi_0\in\left(0, 2\pi\right)$\\
    Polarization & $\Psi\in\left(0, 2\pi\right)$\\
    Inclination & $\cos{\iota}\in\left(-1, 1\right)$\\
    Declination & $\sin{\theta}\in\left(-1, 1\right)$\\
    Right ascension & $\varphi\in\left(-\pi, \pi\right)$\\
    Chirp-Distance & \SI[parse-numbers=false]{d_c^2\in\left(130^2, 350^2\right)}{\mega\parsec^2}\\
    \hline\hline
\end{tabularx}
\caption{A summary of the distributions shared between all datasets from which parameters are drawn.}\label{tab:shared_params}
\end{table}

\subsubsection{Dataset 1}
The noise from the first dataset is purely Gaussian and simulated from the \acrshort{psd} model \verb|aLIGOZeroDetHighPower|~\citep{lalsuite} for both detectors. This means that the \acrshort{psd} used to generate the data contains no sharp peaks originating from factors such as the power grid, is the same for all segments, and is known to the participants.

Injected signals are non-spinning and no higher-order modes are simulated. The component masses are uniformly drawn from \SIrange{10}{50}{M_\odot}. We enforce the condition that the primary mass has to be equal or larger than the secondary mass. With this mass range, at a lower frequency cutoff of \SI{20}{\hertz}, and for non-spinning systems the signal duration is on the order of \SI{1}{\second}.

The first dataset represents a solved problem, as it has already been excessively studied in the past~\citep{George:2016hay, Gabbard:2017lja, Schafer:2021cml}. It is meant as a starting point where people new to the field can refer to existing literature to get off the ground initially. We expected many of the algorithms to perform equally well on this set.

The final challenge set for dataset 1 was generated with the seed $1\,068\,209\,514$ and a start time of $0$.

\subsubsection{Dataset 2}\label{sec:data:ds2}
The noise for the second dataset is also purely Gaussian and simulated. However, in contrast to the first dataset the \acrshort{psd}s were derived from real data from \acrshort{o3a} and as such contain power peaks at certain frequencies and are noisy. We generated a total of $20$ \acrshort{psd}s for each detector. The \acrshort{psd}s used to generate the noise are randomly chosen from these lists and as such are unknown to the search algorithm. The lists themselves are known to the participants. The \acrshort{psd}s in both detectors are independent of each other but do not change over time.

Signals are now allowed to have a spin aligned with the orbital angular momentum with a magnitude between $-0.99$ and $0.99$. Additionally, the mass range is adjusted to draw component masses from the range \SIrange{7}{50}{M_\odot}. This change increases the maximum duration of the signals at a lower frequency cutoff of \SI{20}{\hertz} to $\approx$ \SI{20}{\second}. No higher-order modes are simulated for this dataset and due to the aligned spin requirement no precession effects are present in the waveform.

The second dataset was intended to pose a considerable increase in difficulty to the first dataset. Using an unknown \acrshort{psd} which was derived from real data requires participants to estimate it during the analysis, if the algorithm requires it. However, we expected that increasing the signal duration to up to \SI{20}{\second} would be the more prominent reason for an increase in difficulty as many previous machine learning algorithms have had trouble when dealing with large inputs~\citep{Dreissigacker:2020xfr, Schafer:2020kor, Gabbard:2019rde, Goodfellow:2016aaa}. Finally, we did not expect a large increase in the difficulty of the dataset due to the inclusion of aligned spins.

The final challenge set for dataset 2 was generated with the seed $2\,743\,406\,703$ and a start time of $0$.

\subsubsection{Dataset 3}
The noise for the third dataset is also simulated and purely Gaussian. The increase in difficulty of the noise comes from varying the \acrshort{psd}s over time. Instead of choosing a single random \acrshort{psd} from the list of $20$ \acrshort{psd}s per detector described in \autoref{sec:data:ds2} and generating all noise with that one \acrshort{psd}, the \acrshort{psd} for dataset 3 is randomly chosen for each segment. 

The mass range from \SIrange{7}{50}{M_\odot} and subsequently the maximum signal duration of \SI{20}{\second} is unchanged compared to \autoref{sec:data:ds2}. However, instead of requiring the spins to be aligned with the orbital angular momentum, their orientation is isotropically distributed with a magnitude between $0$ and $0.99$. As a consequence, precession effects are now present in the waveforms. Additionally, we also model all higher-order $(l,m)$-modes available in \verb|IMRPhenomXPHM|, which are: $(2, 2)$, $(2, -2)$, $(2, 1)$, $(2, -1)$, $(3, 3)$, $(3, -3)$, $(3, 2)$, $(3, -2)$, $(4, 4)$, $(4, -4)$~\citep{Pratten:2020ceb}.

The main challenge of this dataset was intended to be the inclusion of precession effects. While these are not as impactful for short duration, high mass systems, they can substantially alter the signal morphology for lower mass systems. Adding higher-order modes can also substantially increase signal complexity. Both of these effects are currently not modeled in any production search relying on accurate signal models, as their inclusion requires an increase in size of the filter bank to include millions of templates~\citep{Harry:2016ijz, Harry:2017weg}. As such, we expected many if not all of the submitted algorithms to struggle with this dataset. On the other hand, any machine learning based algorithm that operates successfully on this dataset may motivate the utilization of machine learning in production searches in the future by extending the searchable parameter space.

The final challenge set for dataset 3 was generated with the seed $470\,182\,217$ and a start time of $0$.

\subsubsection{Dataset 4}
Dataset 4 is the only dataset that contains real detector noise obtained from the Gravitational Wave Open Science Center (\acrshort{gwosc})~\citep{LIGOScientific:2019lzm}. All noise was sampled from parts of \acrshort{o3a} that had the ``data'' quality flag and none of the flags ``CBC\_CAT1'', ``CBC\_CAT2'', ``CBC\_HW\_INJ'', or ``BURST\_HW\_INJ'' were active. We consider only segments where the data from both LIGO Hanford and LIGO Livingston clear the above conditions and excluded \SI{10}{\second} around any detection listed in GWTC-2~\citep{LIGOScientific:2020ibl}. Afterwards we discarded any segments shorter in duration than \SI{2}{\hour}. To allow for different noise realizations, we shift the data from LIGO Livingston by a random time from \SIrange{0}{240}{\second} while keeping the data from LIGO Hanford fixed. The time shifts are independent for each segment and to avoid any possible overlap between neighbouring segments, we consider each segment on its own.

To reduce the amount of data that has to be downloaded by participants we pre-selected the suitable parts of the \acrshort{o3a} data. We then applied a low frequency cutoff of \SI{15}{\hertz} and scaled the data by a factor of $\approx 2^{69}$. Finally, the data was converted to single precision and stored in a compressed format. This allowed us to provide a download link to a single file of \SI{94}{\giga\byte} size containing enough data to generate up to \SI{7024699}{\second}$\approx$\SI{81}{\day} of coincident real noise for both detectors. The data was scaled by the constant factor to avoid the loss of dynamic range due to the conversion from double precision to single precision. When generating test data, the data is converted back to double precision and the scaling is inverted. The code used to downsample the data is also open source and available at~\citep{github}.

The signals are generated equivalently to the signals in dataset 3, i.e. masses are uniformly drawn from \SIrange{7}{50}{M_\odot}, spins are isotropically distributed with a magnitude from $0$ to $0.99$, and all higher-order modes available in \verb|IMRPhenomXPHM| are generated. Consequently, precession effects are simulated.

This dataset is intended to be indicative of a real-world application of the search in parameter regions which are currently sparsely searched. Given that many machine learning searches have proven to generalize well from Gaussian noise to real detector noise at higher \acrshort{far}s in the past~\citep{George:2017pmj, Gebhard:2019ldz, Krastev:2020skk, Wei:2020ztw} we expected that machine learning algorithms that do well on dataset 3 will also be competitive for dataset 4. However, it was expected that handling short glitches may prove difficult for certain searches, especially those focusing most on the merger and ringdown.

The final challenge set for dataset 4 was generated with the seed $2\,514\,409\,456$ and a start time of $0$.

\subsection{Evaluation}\label{sec:metrics}
All submissions are evaluated on the challenge sets, which are generated with a seed unknown to the participants at the time of submission. The evaluation is run on the Atlas computing cluster at the Albert-Einstein-Institut (AEI), Hannover. Groups that submitted an algorithm had no direct access to the evaluation stage\footnote{This excludes submissions by the organization group. However, no member of the organization group accessed the challenge-data before the submission deadline or altered their algorithm after the submission deadline.} and final results presented in this work were only communicated back to the groups after the submission deadline had passed.

We compute two metrics for every submission and dataset. These are the wall-clock time required by the algorithm at hand to analyze one month of data as well as the sensitive distance of the search as a function of the false-alarm rate. In essence, the sensitivity as a function of the false-alarm rate is a receiver operating characteristic (\acrshort{roc}) curve that factors in the varying signal strengths of the injected \acrshort{gw}s. It is a common measure of search sensitivity for production \acrshort{gw}-searches~\citep{Usman:2015kfa} and thus allows for easy comparisons. We do not compute the \acrshort{roc} curve directly, for two reasons. First, it requires the number of a negative samples in the data. Since our data is continuous and the evaluation is left to the groups, defining a negative sample is not possible. Second, the \acrshort{roc} curve can be changed by choosing a different signal population. For instance, the \acrshort{roc} curve can be driven to zero by choosing a population of signals that are excessively far from the detectors. The sensitive distance normalizes the data by the injected population.

For the calculation of the sensitive distances we use two challenge sets for each of the 4 datasets. The first contains pure noise and we will call it the background set from here on out. The second contains the same noise as the background set but adds \acrshort{gw} signals into it. This second set will be called the foreground set from here on out. As described in \autoref{sec:submission-req} any search algorithm is expected to process these files and return lists of events, where an event is a combination of a GPS time, a ranking statistic-like quantity, and a value for the timing accuracy. We will call these events background or foreground events when they have been derived from the background or foreground set, respectively. For the remainder of this section we will refer to the ranking statistic-like quantity simply as ranking statistic, to simplify our statements.

To calculate the sensitivity as a function of the false-alarm rate, we need to determine the false-alarm rate as a function of the ranking statistic. Next we can also determine the sensitivity as a function of the ranking statistic. Finally, we can combine the two, by evaluating both at the same values of the ranking statistic.

We use the ranking statistic of all background events as points where both the \acrshort{far} as well as the sensitivity is evaluated. Each of these is certain to be a false positive and thus ensures that the \acrshort{far} is unique at each threshold, as long as the search does not return identical ranking statistics for multiple background events.

To calculate the \acrshort{far} at a given ranking statistic we count the number of background events with a ranking statistic greater than this threshold. We, subsequently, turn that into a rate by dividing the number of false-positives by the duration of the background data, i.e. \SI{2592000}{\second}. With $N_{\text{FP}, \mathcal{R}}$ the number of false-positives at a given ranking statistic $\mathcal{R}$ and $T$ the time spanned by the background set, the \acrshort{far} $\mathcal{F}$ can be calculated by
\begin{equation}
    \mathcal{F} = \frac{N_{\text{FP}, \mathcal{R}}}{T}.
\end{equation}

The sensitive volume of a search at \acrshort{far} $\mathcal{F}$ can be calculated by~\citep{Usman:2015kfa}
\begin{equation}\label{eq:sens-vol-full}
    V\left(\mathcal{F}\right) = \int d\bm{x}d\bm{\Lambda}\ \epsilon\left(\mathcal{F};\bm{x},\bm{\Lambda}\right)\phi\left(\bm{x},\bm{\Lambda}\right),
\end{equation}
where $\bm{x}$ are the spatial coordinates of the injection, $\bm{\Lambda}$ are the injection parameters, $\epsilon\left(\mathcal{F};\bm{x},\bm{\Lambda}\right)$ is the efficiency of the search at \acrshort{far} $\mathcal{F}$, and $\phi\left(\bm{x},\bm{\Lambda}\right)$ is the distribution of the injection parameters $\bm{x}$ and $\bm{\Lambda}$.

When injections are performed uniformly in volume up to a maximum distance $d_\text{max}$, \autoref{eq:sens-vol-full} can be approximated by~\citep{Usman:2015kfa}
\begin{equation}\label{eq:sens-vol-approx-1}
    V\left(\mathcal{F}\right)\approx V\left(d_\text{max}\right)\frac{N_{I,\mathcal{F}}}{N_I},
\end{equation}
where $ V\left(d_\text{max}\right)$ is the volume of a sphere with radius $d_\text{max}$, $N_{I,\mathcal{F}}$ is the number of found injections at a \acrshort{far} of $\mathcal{F}$, and $N_I$ is the total number of injections performed. An injection is found if there is at least one foreground event that is within $\pm \Delta t$ of the injection, where $\Delta t$ is the time variance assigned to the event by the search algorithm. The number of found injections at a given \acrshort{far} considers only those foreground events where the ranking statistic assigned to the specified event is greater than the ranking statistic corresponding to the \acrshort{far}. In machine learning terms \autoref{eq:sens-vol-approx-1} is the recall at a given threshold on the network output multiplied by the volume of a sphere with radius $d_\text{max}$, assuming that each injection corresponds to exactly one true positive.

However, the injections in the datasets are not performed uniformly in volume, as we sample over the chirp-distance instead of the luminosity distance. The chirp-distance is given by~\citep{LIGOScientific:2007npa}
\begin{equation}
    d_c = d{\left(\frac{\mathcal{M}_{c,0}}{\mathcal{M}_c}\right)}^{5/6},
\end{equation}
where $d$ is the luminosity distance, $\mathcal{M}_c = {\left(m_1 m_2\right)}^{3/5} / {\left(m_1 + m_2\right)}^{1/5}$ is the chirp-mass, and $\mathcal{M}_{c,0} = $\SI[parse-numbers=false]{1.4/2^{1/5}}{M_\odot} is a fiducial chirp-mass used as a basis for calculation. Note that in contrast to \citep{LIGOScientific:2007npa} we use the luminosity distance instead of the effective distance as our basis.

When sampling the injections from the distributions defined in \autoref{tab:shared_params} using the chirp-distance, effectively the maximum luminosity distance $d$ is selected based on the chirp-mass; the smaller the chirp mass, the smaller the maximum luminosity distance at which injections are placed. This allows us to increase the number of detectable low mass systems and, subsequently, make statistically meaningful statements about the sensitivity for these systems without requiring a large increase in the amount of data that needs to be analyzed. However, when considering a fixed chirp mass, injections are still placed uniformly within that sphere of the adjusted maximum luminosity distance. In \autoref{eq:sens-vol-approx-1} we assumed that each injection was placed uniformly within the volume spanned by the sphere with volume $V\left(d_\text{max}\right)$. To adjust it for sampling over luminosity distance we have to factor in that the probed distance depends on the selected chirp mass. We, therefore, find
\begin{equation}
     V\left(\mathcal{F}\right)\approx \frac{V\left(d_\text{max}\right)}{N_I}\sum_{i=1}^{N_{I,\mathcal{F}}} \frac{V\left(d_{c,\text{max}} {\left(\frac{\mathcal{M}_{c, i}}{\mathcal{M}_{c,0}}\right)}^{5/6}\right)}{V\left(d_{c,\text{max}} {\left(\frac{\mathcal{M}_{c, \text{max}}}{\mathcal{M}_{c,0}}\right)}^{5/6}\right)},
\end{equation}
where $\mathcal{M}_{c,i}$ is the chirp mass of the $i$-th found injection, $d_{c,\text{max}}$ is the upper limit on the injected chirp distances, and $\mathcal{M}_{c,\text{max}}$ is the upper limit on the injected chirp masses. This expression can be simplified to yield
\begin{equation}
     V\left(\mathcal{F}\right)\approx \frac{V\left(d_\text{max}\right)}{N_I}\sum_{i=1}^{N_{I,\mathcal{F}}} {\left(\frac{\mathcal{M}_{c, i}}{\mathcal{M}_{c, \text{max}}}\right)}^{5/2},
\end{equation}
which is the formula we use to estimate the sensitive volume of a search algorithm. Instead of quoting the volume directly we convert it to the radius of a sphere with the corresponding volume and quote that instead.

We also measure the time the algorithm requires to evaluate an entire month of test data. Since all machine learning search algorithms are running on the same hardware these values can be used to compare the speed of the different analyses on the given hardware. For a summary of the available hardware resources please refer to \autoref{tab:hardware}. However, we expect the computational time to be dominated by pre-processing steps, which can in theory be heavily optimized. For this challenge, though, we did not expect many submissions to invest resources into optimizing their pre-processing and thus advise the reader to not overemphasize the provided numbers.

\begin{table}[]
    \centering
    \begin{tabularx}{0.48\textwidth}{XX}
    \hline\hline
    Hardware type & Specification \\
    \hline
    CPU & $2\times$ Intel Xeon Silver 4215, 8(16) cores(threads) at \SI{2.5}{\giga\hertz} \\
    GPU & $8\times$ NVIDIA RTX 2070 Super (\SI{8}{\giga\byte} VRAM) \\
    RAM & \SI{192}{\giga\byte} \\
    \hline\hline
    \end{tabularx}
    \caption{Main hardware specifications available to each search algorithm during final testing.}
    \label{tab:hardware}
\end{table}

All runtimes are measured twice; once for the foreground set and once for the background set. In both cases the wall-time that has passed between calling the executable and it returning is measured.

\subsection{Submission Requirements}\label{sec:submission-req}
All submissions are provided with the path to a single file containing the input data they have to process. In particular they have to be able to read HDF5 files, the structure of which is detailed in \autoref{sec:data}. Importantly, no pre-processing other than the introduction of a low frequency cutoff of \SI{15}{\hertz} has been applied to the data. All other pre-processing has to be performed by the algorithms themselves. In addition to the path to the input data, each algorithm is provided with a second path at which it is expected to store a single HDF5 file. This file has to contain three one-dimensional datasets of equal size named ``time'', ``stat'', and ``var''.

The ``time'' dataset is expected to contain the GPS times at which the algorithm predicts a \acrshort{gw} signal to be present. These are compared to the injection times to determine which injections were found, which were missed, and how many false positives the analysis produced.

The ``stat'' dataset is expected to contain a ranking-statistic like quantity for every GPS time in the ``time'' dataset. Here, ranking-statistic like quantity means a value where larger numbers indicate a higher degree of believe for the search to have found a \acrshort{gw} signal. Having a ranking-statistic like quantity associated to all candidate detections enables us to assign a statistical significance to any event.

The ``var'' dataset is expected to contain the estimated timing accuracy of the search algorithm for all GPS times in the ``time'' dataset. This value determines the window around the GPS time returned by the search within which an injection has had to be made in order to consider the detection a true positive and the injection to be found. This value may be constant for all times at which the search expects to have seen a signal. We allowed searches to specify this value themselves, as we felt it to be unsuitable for a signal detection challenge to require a fixed timing accuracy. In principle, this freedom can be abused by choosing an accessively high value of $\Delta t$ and claiming all events as true positives. However, all groups have chosen values on similar scales and more importantly far shorter than the average separation of two injections.

Throughout the paper, we will refer to events returned by the search. By that we mean a single tuple $\left(t, \mathcal{R}, \Delta t\right)$ contained in the ``time'', ``stat'', and ``var'' datasets, respectively.

To be able to execute all algorithms without major problems, we ask participants to either provide a single executable that can be run on the Linux command-line utilizing only the provided software stack or to provide a singularity image that we can execute. In both cases the algorithms have to accept two positional command line arguments; the path to the input data file and the path at which the output file should be stored. The main Python packages available to submitted executables are listed in \autoref{tab:software}, for a full list refer to \citep{github}.

\begin{table}[]
    \centering
    \begin{tabularx}{0.48\textwidth}{XX}
        \hline\hline
        Python package & Version \\
        \hline
        bilby & 1.1.3 \\
        pycbc & efeaeb6 \\
        tensorflow-gpu & 2.6.0 \\
        tensorflow-probability & 0.14.0 \\
        torch & 1.9.1+cu11 \\
        \hline\hline
    \end{tabularx}
    \caption{A selective list of the core Python packages available to algorithms during evaluation. A complete list is given at~\citep{github}.}
    \label{tab:software}
\end{table}

Each algorithm is executed by hand and closely monitored by the organization team of the challenge. Participants are not allowed to directly tune or influence the final evaluation.

To ensure that participants have submitted the correct version of their algorithm and to make sure that their algorithm behaves as expected on the evaluation hardware and software, all algorithms are first evaluated on a validation set which is generated equivalently to the final test set. The results on this validation set are then communicated back to the submitting group. Once the group has approved that their algorithm performs within the expected margin of error, the algorithm is applied to the real challenge sets. These challenge sets are the same for all participants and were kept secret until the deadline for final submissions had passed.

Since multiple members of the organization team have submitted algorithms to this challenge, the challenge datasets were only generated after the submission deadline had passed. The script to generate test data provides an option to use a random seed. This option was used to generate the final challenge datasets and ensures that no submission had knowledge of the challenge set prior to the submission deadline.

We allowed all participants to retract their submissions at any point prior to the final publication of our results. This means that participants were allowed to retract their submissions even after they were informed about the performance of their algorithm on the final challenge sets and after they have seen the performance of other entries. No group made use of this freedom and retracted their submission after results were internally published.

\section{Submissions}\label{sec:submissions}
In this section we briefly introduce the different algorithms. For more details on the individual submissions we refer the reader to the original works cited within each subsection. The subsections are titled by the group name and are given in order of registration to the challenge.

All algorithm preparation was performed by the individual groups using their own available hardware resources. This crucially includes training of machine learning algorithms, for which no resources were provided by the organizers of this challenge. There were no strict requirements to submit algorithms that are based on machine learning techniques. We even encouraged the submission of a few traditional algorithms to quote a point of reference. However, the available resources detailed in \autoref{sec:metrics} for evaluation of the test sets are tailored to suit the needs of machine learning algorithms.

\subsection{MFCNN}\label{sec:submission-mfcnn}
\footnote{The corresponding authors for the MFCNN submission are He Wang, Shichao Wu, Zong-Kuan Guo, Zhoujian Cao, and Zhixiang Ren}
The submission of the MFCNN group is based on the works from He et al.~\citep{Wang:2019zaj}. The authors of \citep{Wang:2019zaj} refer to the model as matched-filtering convolutional neural network (MFCNN). MFCNN is a semi-coherent search model. The basic idea of the model is to use waveform templates as learnable weights in neural network layers. Analogously to the standard coincident matched-filtering searches the output of each matched-filtering layer is maximized and normalized in the unit of matched-filtering \acrshort{snr}s for each \acrshort{gw} detector. However, triggers are not generated on a single detector. The remaining part of the neural network is a usual convolutional neural network that is employed afterwards to jointly analyze the output from all detectors. Finally, a SoftMax function is applied to evaluate the confidence score of a \acrshort{gw} signal being present in the \acrshort{gw} detector network. The architecture was designed to take the advantages of both matched-filtering and convolutional neural networks and combine them to search for real \acrshort{gw} events in GWTC-1~\citep{LIGOScientific:2018mvr}. To adapt to this challenge, the source code~\citep{Wang:2022aaa} of the submission was translated from the MXNet framework~\citep{Chen:2015aaa} used in the original work to a PyTorch~\citep{pytorch_software} implementation.

The training data for the model is generated by the code that generates dataset 4. The training data are input into the model directly with none of the usual pre-processing such as band-pass or whitening, which is consistent with the original work~\citep{Wang:2019zaj}. In fact, the model is equipped with a whitening layer to estimate the power spectrum for each input data. The main modification used in this challenge is to randomly sample 25 templates in the first matched-filtering layer from the same parameter space used in dataset 4 of this challenge. It performs significantly better than the original gridded and fixed template configuration. The subsequent convolution network of the model is constructed using the current excellent lightweight models MobileNetV3~\citep{Howard:2019aaa} which give state-of-the-art results in major computer vision problems. The submission uses curriculum learning, during which the model is trained with decreasing multiples of signal amplitude. The multiplicative factor is lowered from 50 to 1 until convergence. Multiple models were randomly initialized and trained on a NVIDIA Tesla V100 GPU, from which the best was chosen for this submission.

To search for triggers and evaluate the performance of the model, a sliding window approach is implemented. The evaluation data is divided into overlapping segments corresponding to the input size of the model. Subsequently, all segments are passed through the model resulting in a sequence of predictions and a table of \acrshort{snr} peaks from the 25 sorted matched filters. The step size is 1 second and a threshold of 0.5 is set on the network output as in~\citep{Wang:2019zaj}. The ``time''-, ``var''- and ``stat''- dataset of the output file described in \autoref{sec:submission-req} are derived from the table of \acrshort{snr} peaks associated with directly filtering the templates with the data. The GPS time and time variance of each trigger are designated as the median value and the interquartile range of \acrshort{snr} peaks from the nearby segments, respectively. We count the coincident \acrshort{snr} peaks between two detectors to quantify the ranking-statistic. Other experiments are still in progress and are supposed to be published alongside further details in a standalone paper.

The final version of the algorithm submitted by the \mfcnn group was provided after the submission deadline had past. A vital flaw in their original contribution was discovered and was allowed to be fixed.

\subsection{PyCBC}\label{sec:submission-pycbc}
\footnote{The corresponding author for the PyCBC submission is Alexander H. Nitz.}
The \textit{PyCBC} submission is based on a standard
configuration of the PyCBC-based archival search 
for compact-binary mergers~\citep{Nitz:2021zwj}. The search infrastructure was used, in addition to cWB, for the first detection of gravitational waves, GW159014~\citep{LIGOScientific:2016aoc}, in production analyses by multiple groups to produce gravitational-wave catalogs~\citep{LIGOScientific:2021djp, Nitz:2021zwj} and targeted analyses~\citep{Nitz:2019spj}. A similar low-latency PyCBC-Live analysis is also based around the same toolkit~\citep{Nitz:2018rgo,DalCanton:2020vpm}. The analysis uses matched filtering to identify candidate observations in combination with a bank of predetermined waveform templates that correspond to the expected gravitational-wave signals~\citep{Allen:2005fk}. Matched filtering is known to be the optimal linear filter for stationary, Gaussian noise. To account for the potential non-Gaussian noise transients~\citep{Nuttall:2015dqa,Cabero:2019orq,Davis:2022dnd}, each candidate and the surrounding data are checked for consistency with the expected signal~\citep{Allen:2004gu,Nitz:2017lco}. In addition, the properties of candidates, such as their time of arrival, amplitude, and phases in each detector are checked for consistency with an astrophysical population~\citep{Nitz:2017svb}. 

The empirically measured noise distribution and the consistency with the expected gravitational-wave signal are combined to calculate a ranking statistic for each potential candidate~\citep{Nitz:2017svb, Davies:2020}; this ranking statistic is used as the ``stat'' value of dataset output, along with its associate trigger time in ``time''. The ``var'' dataset is set to a constant of \SI{0.25}{\second}. Two template banks
are used for the submitted results. For dataset 1, a template bank of non-spinning
waveform templates, using the IMRPhenomD~\citep{Husa:2015iqa} model, is created using stochastic placement. Datasets 2, 3, and 4 were evaluated with a common template bank that includes templates that account for spin which is aligned with the orbital angular momentum. Furthermore, only the dominant mode of the gravitational-wave signal was used and effects such as precession were not accounted for. In both cases, the mass boundaries of the template bank conform to the challenge set parameters.

The final version of the algorithm submitted by the \pycbc group was provided after the submission deadline had past. A vital flaw in their original contribution was discovered and was allowed to be fixed. Furthermore, the \pycbc submission strictly speaking uses a different algorithm for dataset 1 than for all other datasets, as the template banks are not the same. The change in template banks was accepted, as this work does not focus on a runtime analysis.

\subsection{CNN-Coinc}\label{sec:submission-cnn}
\footnote{The corresponding author for the CNN-Coinc submission is Marlin B. Sch{\"a}fer.}
This submission is based on the works from Gabbard et al.~\citep{Gabbard:2017lja} and Sch\"afer et al.~\citep{Schafer:2021cml}. It utilizes the network architecture presented in~\citep{Gabbard:2017lja} with a prepended batch-normalization layer~\citep{Ioffe:2015aaa}. As such the network processes $8\,192$ input samples, which corresponds to \SI{4}{\second} at a sampling rate of \SI{2}{\kilo\hertz}. The network is trained only once and applied to the data from both detectors individually. Afterwards the outputs are correlated to find coincident events as detailed in~\citep{Schafer:2021cml}. The source code for training the network and applying it to test data of the format used in this challenge is open source and can be found at~\citep{github_cnn_coinc}. The algorithm was designed to enable an easy and efficient estimation of the search background by applying time shifts between the individual detectors data. While this feature cannot be utilized in this challenge, the original paper~\citep{Schafer:2021cml} highlights the advantages of this approach.

The network is trained on parts of the real \acrshort{o3a} noise from the Hanford detector as provided in this challenge. Signals are generated using the waveform approximant \verb|IMRPhenomXPHM|~\citep{Pratten:2020ceb} from the same parameter distribution used in datasets 3 and 4 in this challenge. Merger times of the signals are varied between \SIrange{2.9}{3.1}{\second} from the start of the input window of the network. The signals are pre-whitened by one of the provided Hanford \acrshort{psd}s used in datasets 2 and 3. Noise samples are non-overlapping parts taken from the real noise data provided by this challenge, where each segment is whitenened by an estimate of the \acrshort{psd} on that segment. The network was trained for 100 epochs using the loss and optimizer settings provided in~\citep{Schafer:2021cml} on a single NVIDIA RTX 2070. The epoch with the greatest binary accuracy on a single training run was chosen for this challenge.

During evaluation the network is applied to the challenge-data using a sliding window approach. Each data segment is whitened by an estimate of the \acrshort{psd} of that segment obtained by Welch's method~\citep{Allen:2005fk, Welch:1967aaa}. All data is whitened before the network is applied for computational efficiency. Subsequently, the network is applied to the data via a sliding window with a step size of $204$ samples \SI[parse-numbers=false]{\approx 0.1}{\second}. Afterwards a threshold of $3.4$ is applied on the unbounded Softmax replacement (\acrshort{usr}) output, which was introduced in~\citep{Schafer:2021fea}. Coincident events are calculated using the same procedure and parameters as outlined in~\citep{Schafer:2021cml}. The ``time''- and ``stat''-dataset of the output file described in \autoref{sec:submission-req} list the coincident event times and ranking statistic values, respectively. The time variance of the ``var''-dataset is set to a constant value of \SI{0.3}{\second}.

\subsection{TPI FSU Jena}\label{sec:submission-jena}
\footnote{The corresponding authors for the \jena submission are Ond{\v{r}}ej Zelenka, Bernd Br{\"u}gmann, and Frank Ohme.}
This submission closely followed the method of \citep{Schafer:2021fea}, which is itself based on~\citep{Gabbard:2017lja}, with several modifications to adapt to the specifics of the challenge. The core of the algorithm is a convolutional neural network that accepts a $2\times 2048$ input tensor corresponding to 1 second of data from 2 detectors sampled at \SI{2048}{\hertz}. Its architecture is derived from that of \citep{Schafer:2021fea} and deviates from the original network by a larger size of the individual layers and a doubled number of convolutional layers. These modifications are the result of a hyperparameter variation experiment which found these settings to be optimal. A standalone publication on this submission giving further details on the methodology is in preparation. The final layer of the network is a Softmax layer over two inputs which is used for training and removed using the \acrshort{usr} \citep{Schafer:2021fea} during evaluation.

The network is trained on a dataset constructed by whitening a randomly chosen part of the real noise file and slicing it to produce 1-second noise samples and injecting whitened \texttt{IMRPhenomXPHM}-generated BBH waveforms into half the noise samples at \acrshort{snr}s uniformly drawn between $7$ and $20$. The waveform parameters are drawn from the same distributions as are used in dataset 4 of this challenge. The training dataset consists of $10^6$ samples and the validation set of $2\cdot 10^5$ samples.

During evaluation, each segment in the input file is whitened separately using the estimated \acrshort{psd} and sliced into 1-second segments at 0.1-second spacing. These are fed to the network with the \acrshort{usr} applied. First-level triggers are selected by applying a threshold of -8, which are then clustered into events. For each event, the ``time'' and ``stat'' in the output file are the values of the highest ranking statistic first-level trigger of each cluster, and ``var'' is set to $0.2$ seconds. The algorithm is implemented using the PyTorch framework~\citep{pytorch_software} and spawns child processes to whiten individual segments. The network evaluation is performed by the parent process.

\subsection{Virgo-AUTh}\label{sec:submission-auth}
\footnote{The corresponding authors for the \virgo submission are Paraskevi Nousi, Nikolaos Stergioulas, Panagiotis Iosif, Alexandra E. Koloniari, Anastasios Tefas, and Nikolaos Passalis.}
This submission is based on a simple per-dataset binary classification scheme. Interestingly, it was found that training a model only on dataset 2 or only on dataset 4 can yield impressive results on the other datasets as well. Specifically, training samples from dataset 2 can generalize well to dataset 3 and 1 and not so well on dataset 4, whereas training samples from dataset 4 can generalize well on datasets 1, 2 and 3. Thus, training samples were only generated from dataset 4. An adaptive normalization mechanism \citep{passalis2019deep} was used instead of batch normalization as the first layer, to handle non-stationary timeseries. For the neural network architecture a deep, ResNet-like model \citep{he2016deep} with a depth of 54 layers was used.

One week of training data per dataset was generated and the generated injection parameters were used to construct all corresponding waveforms. This amounted to about 600k background segments of duration \SI{1.25}{\second} with a stride of \SI{2}{\second} between, i.e. the next sample starts \SI{0.75}{\second} after the end of the previous one, and about 580k waveforms, of which 300k were used for the injections. For validation, one day of data was used, resulting in about 86k noise segments and 3.2k waveforms. The noise segments and waveform segments are combined online during training, in a static manner, both for the training and for the validation sets. The input samples are whitened before feeding them to the classifier. The \acrshort{psd} is computed online per batch of \SI{4.25}{\second} with a stride of \SI{3.1}{\second}, and each \SI{1.25}{\second} segment inside this duration is whitened with the same \acrshort{psd}. To increase speed, the Welch method for computing the \acrshort{psd} was implemented in PyTorch~\citep{pytorch_software} and whitening is implemented as the first layer of the final detection module. Notably, this approach of computing the \acrshort{psd} for every \SI{4.25}{\second} and whitening each \SI{1.25}{\second} segment in a sliding window manner was found to be faster than using a precomputed \acrshort{psd} for every \SI{1.25}{\second} (about $40\%$ faster for one month of data). After whitening, the first and last \SI{0.125}{\second} (\SI{0.25}{\second} total) are removed from each sample.

The best results were obtained with a ResNet-52 type network. A Deep Adaptive Input Normalization (DAIN) layer \citep{passalis2019deep} was used as the first layer after whitening, to handle distribution shifts that may be present. The final output is binary, i.e., noise plus waveforms or noise only, and the objective function used was a regularized binary cross entropy. The ``var'' parameter is set to \SI{0.3}{\second}, as the network predictions are high even when the time of coalescence is slightly outside the preset range. The ``stat'' parameter is set to the network confidence, i.e., a value in the $[0, 1]$ interval corresponding to the probability that a waveform is present. Finally, \SI{0.125}{\second} are added to the expected time of coalescence to account for the time lost in the whitening process.

A standalone publication on the methods used in this submission is in preparation.

\subsection{cWB}\label{sec:submission-cwb}
\footnote{The corresponding authors for the \cwb submission are Francesco Salemi, Gabriele Vedovato, Sergey Klimenko, Tanmaya Mishra.}
Coherent WaveBurst (\cwb) is a waveform model-agnostic search pipeline for \acrshort{gw} signals based on the constrained likelihood method \citep{SoftX, Zenodo, cWB_homepage}. The \cwb pipeline has been used for the analysis of scientific data collected by the LIGO-Virgo detectors,  targeting detection of signals from generic \acrshort{gw} sources, including the compact binary mergers~\citep{LIGOScientific:2021djp}.

The \cwb algorithm identifies the excess-power events in the time-frequency domain representation of strain data from multiple detectors~\citep{Klimenko:2015ypf,CWBwavelet:2004km}. For each event, the \cwb pipeline reconstructs the \acrshort{gw} waveforms and estimates summary statistics which describe generic properties of the events like the coherence across the detector network, signal strength, and the time-frequency structure.
 
Recently, a boosted decision-tree algorithm, eXtreme-Gradient Boost (XGBoost)~\citep{XGBoost}, was adopted and implemented within the \cwb framework to automate the signal-noise classification of the \cwb events~\citep{Mishra:2021tmu}. Two types of input data are used for the supervised training: signal events (from simulations) and noise events (from background estimations). For each of those, a subset of \cwb summary statistics is fed to XGBoost as input features to train a signal-noise model.  
As in ~\citep{Mishra:2021tmu}, the detection statistic for the machine learning-enhanced \cwb algorithm is defined by:
\begin{equation}\label{eq:x}
	\eta_\mathrm{r} = \eta_\mathrm{0}\cdot W_{\mathrm{XGB}}, 
\end{equation}

where, $\eta_\mathrm{0}$ is {\cwb}s ranking statistic, and $W_{\mathrm{XGB}}$ is the penalty factor calculated by XGBoost ranging between 0 (noise) and 1 (signal). 

This methodology has been recently used in the full reanalysis of publicly available strain data from Advanced LIGO’s Hanford and Livingston third observational run~\citep{Mishra:2022ott}: the machine learning-enhanced \cwb outperforms the standard human-tuned signal-noise classification used for detection of the compact binary coalescences in the \acrshort{o3} run.

For this study, we chose to use machine learning-enhanced \cwb; however, \cwb typically rejects weak candidate triggers (i.e., with \acrshort{far} $\gg 1$ per year) at early production stages. Moreover, the whole workflow is optimized for a trigger production which saturates at \acrshort{far}~$\approx 30 \text{ to } 50$ per month.  Therefore, we modified \cwb to increase the event production rate by almost 2 orders of magnitude: the result is a \cwb with sub-threshold capabilities, able to speed up computation and reduce memory allocations.

While trying to provide the most ``generic'' result for this study, it was decided to re-use the XGBoost model which was developed for ~\citep{Mishra:2022ott}: it should be noted that the model was trained on noise and signal events sets that differ substantially from those adopted for the data sets prepared for \acrshort{mlgwsc1}. The noise backgrounds for dataset 3 and dataset 4 appear to be significantly quieter than \acrshort{o3}. Also, the signals were drawn from a spin-aligned stellar-mass \acrshort{bbh}s population model with different component mass ranges~\citep{Power_law_peak} and with SEOBNRv4 waveforms ~\citep{Boh2017}.
The above-mentioned detection statistic, $\eta_\mathrm{r}$, is used as the ``stat'' value of dataset output, along with its associated trigger peak-time in ``time''. The ``var'' dataset is set to a constant of \SI{0.25}{\second}.

The results from the \cwb group were provided after the submission deadline had passed. The group assured that no tuning to the challenge set was performed.

\section{Data release}
We provide all source code as well as the evaluation results for all submissions at \citep{github}. The repository contains all code accessible to the participants of the challenge, which most importantly  includes a script to generate data and one to produce the sensitivity statistics we provide in \autoref{sec:results}. The repository also contains code for basic visualization as part of the ``contributions'' folder. Adaption of these scripts were used to create the graphics in this paper. The challenge used the code of release 1.3 of the repository.

Alongside the code provided by the challenge organizers we publish the source code that was used to run the contributions for the groups \pycbc, CNN-Coinc, \jena, and \virgo in the ``submissions'' folder of \citep{github}. The submission code for the MFCNN group can be found at \citep{Wang:2022aaa}.

All analysis output files for all submissions created by our analysis are also publicly available and are stored in the ``results'' folder in \citep{github}. For each group we make available the raw output on the foreground and the background for all 4 datasets. Additionally, all timing information is available. The exception is the \cwb group, for which only results on datasets 3 and 4 are available. 

The repository \citep{github} also contains plots used in this paper for all groups, including versions we have not shown here. They can be found in the ``plots'' folder.

\section{Results and discussion}\label{sec:results}
In this section we provide the results of our evaluation process described in \autoref{sec:methods} for all $6$ submissions. We calculate and discuss sensitive distances, found-missed plots, and runtimes to provide a quantitative comparison between the different submissions. We specifically focus on the difference between machine learning and traditional algorithms and reason where the core differences in performance arise.

The four datasets we use in this study were chosen to answer different questions and serve different purposes. Dataset 1 was meant as an entry point to the challenge that represents a largely solved case~\citep{George:2016hay, Gabbard:2017lja, Schafer:2021cml}. We expected most submissions to perform very similarly on this dataset. The second dataset was intended as the first major step in difficulty. We expected its main challenge to be the longer duration of the injected signals, as many machine learning algorithms target shorter durations and struggle with large analysis segments~\citep{Goodfellow:2016aaa, Schafer:2020kor}. Dataset 3 includes precession and higher order mode effects in the injected signals that traditional, modeled searches are not optimized for\footnote{A full search of the entire \acrshort{o3} data that includes higher order modes has been performed in \citep{Chandra:2022ixv}.}~\citep{Harry:2016ijz, Harry:2017weg, Dhurkunde:2022aek}. We wanted to test if machine learning algorithms could get closer in performance, or even outperform, the traditional searches in these regions. The intention of dataset 4 was to provide a challenge that is representative of a realistic search on real detector data and a limited parameter space. The data contains non-Gaussian noise artifacts, that can mimic \acrshort{gw} signals~\citep{LIGOScientific:2014qfs, LIGOScientific:2016gtq, LIGOScientific:2019hgc, Davis:2022cmw}, which are strongly suppressed by sophisticated algorithms in traditional searches~\citep{Usman:2015kfa, Messick:2016aqy, LIGOScientific:2019hgc}. Most machine learning algorithms that target real noise do not make use of such noise-mitigation strategies and instead rely solely on the ability of the machine learning algorithm to identify noise artifacts. This approach was reported to be effective for higher \acrshort{far}s in the past~\citep{George:2017pmj, Gebhard:2019ldz, Krastev:2020skk, Wei:2020ztw} and we were, therefore, expecting relatively minor difference between dataset 3 and dataset 4. Furthermore, most traditional algorithms use matched filtering, which is only proven to be optimal for signal recovery when the noise is stationary and Gaussian. Since neither of the two assumptions are true for real detector data, we were also interested to test if machine learning algorithms can perform better than these searches by learning a better noise representation.

\subsection{Sensitivities}
In this subsection we discuss the sensitive distances of the different submissions, which are a measure for how many sources can be detected at any given level of certainty, i.e. at a particular \acrshort{far}. They are the core metric to determine the quality of any search. We focus on the low \acrshort{far} region and truncate the plot at a \acrshort{far} of $10^3$ per month. We chose this cutoff for two reasons. First, to function as a standalone search, algorithms may only report events with low \acrshort{far}s. State of the art pipelines send out alerts only when the \acrshort{far} is smaller than $\mathcal{O}\left(1\right)$ per month~\citep{Nitz:2018rgo}. Second, for high \acrshort{far}s a non-negligible number of detections originate from false associations. This means that a large number of triggers that originate from random noise coincidences are close enough to an injection to be counted as true positives.

Since all machine learning submissions chose to optimize for dataset 4, results on all prior sets also test the capability of generalizing to different signal (sub)populations. Dataset 3 is a special case, as it uses the same distribution to draw the parameters of the injected signals as dataset 4. It, therefore, differs only in the noise contents and is a good test of the performance difference of different algorithms between simulated and real noise.

\begin{figure*}
    \centering
    \includegraphics[width=0.98\textwidth]{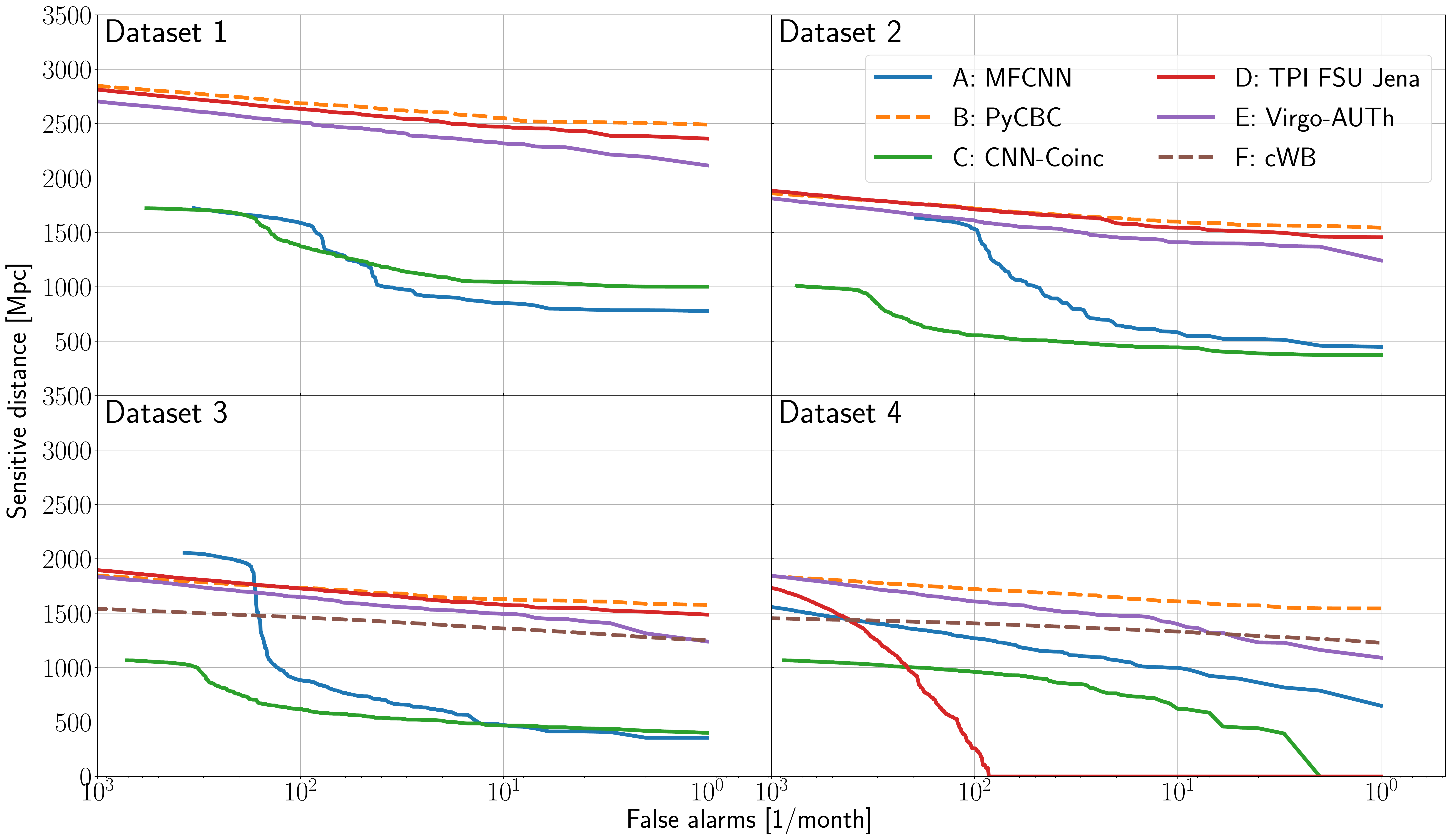}
    \caption{The sensitive distances of all submissions and all four datasets as functions of the \acrshort{far}. Submissions that made use of a machine learning algorithm at their core are shown with solid lines, others with dashed lines. The \acrshort{far} was calculated on a background set that does not contain any injections.}
    \label{fig:sens}
\end{figure*}

The results of this challenge are summarized in \autoref{fig:sens} and \autoref{tab:res}. The four individual panels of \autoref{fig:sens} show the sensitive distances as a function of the \acrshort{far} for all submissions. The panels contain the results for dataset 1 to 4 from left to right and top to bottom. In \autoref{tab:res} we give the numeric values for the sensitive distances at three selected \acrshort{far} values of $1$, $10$, and $100$ per month for all submissions and datasets. We also provide information on the wall-clock time used to evaluate the different sets. Due to time constraints, we only show sensitivity curves for dataset 3 and 4 for the submission from the \cwb group. We also note that \pycbc used a different template bank to analyze dataset 1 than for the remaining three datasets.

\begin{table*}[]
    \centering
    \begin{tabularx}{0.98\textwidth}{|X|X|X|X|X|X|X|X|}
    \hline
    \multirow{2}{*}{Dataset} & \multirow{2}{*}{Group} & \multicolumn{3}{c|}{Sensitivity [Mpc] at \acrshort{far} $=x$ per month} & \multicolumn{3}{c|}{Runtime [s]} \\
    \cline{3-8}
     & & $x=100$ & $x=10$ & $x=1$ & foreground & background & average \\
    \hline
    \cellcolor{Gray} &  A: MFCNN &  1586.90 &  852.18 &  779.21 &  42842 &  43820 &  43331 \\
    \cellcolor{Gray} & \cellcolor{Gray}B: PyCBC & \cellcolor{Gray}2686.55 & \cellcolor{Gray}2550.57 & \cellcolor{Gray}2491.53 & \cellcolor{Gray}5406$^\ast$ & \cellcolor{Gray}5092$^\ast$ & \cellcolor{Gray}5249$^\ast$ \\
    \cellcolor{Gray} &  C: CNN-Coinc &  1372.30 &  1045.34 &  1001.55 &  14003 &  12996 &  13500 \\
    \cellcolor{Gray} & \cellcolor{Gray}D: TPI FSU Jena & \cellcolor{Gray}2634.80 & \cellcolor{Gray}2472.31 & \cellcolor{Gray}2362.51 & \cellcolor{Gray}3758 & \cellcolor{Gray}3530 & \cellcolor{Gray}3644 \\
    \cellcolor{Gray} &  E: Virgo-AUTh &  2511.95 &  2317.53 &  2116.38 &  5490 &  5520 &  5505 \\
    \multirow{-6}{*}{\cellcolor{Gray}1} & \cellcolor{Gray}F: cWB & \cellcolor{Gray}N/A & \cellcolor{Gray}N/A & \cellcolor{Gray}N/A & \cellcolor{Gray}N/A & \cellcolor{Gray}N/A & \cellcolor{Gray}N/A \\
    \hline
      &  A: MFCNN &  1531.55 &  581.93 &  448.59 &  43431 &  40634 &  42033 \\
      & \cellcolor{Gray}B: PyCBC & \cellcolor{Gray}1719.98 & \cellcolor{Gray}1599.79 & \cellcolor{Gray}1543.79 & \cellcolor{Gray}157865$^{\ast\ast}$ & \cellcolor{Gray}161662$^{\ast\ast}$ & \cellcolor{Gray}159763$^{\ast\ast}$ \\
      &  C: CNN-Coinc &  554.32 &  443.58 &  373.64 &  14731 &  14976 &  14853 \\
      & \cellcolor{Gray}D: TPI FSU Jena & \cellcolor{Gray}1712.13 & \cellcolor{Gray}1544.28 & \cellcolor{Gray}1455.09 & \cellcolor{Gray}3920 & \cellcolor{Gray}3805 & \cellcolor{Gray}3862 \\
      &  E: Virgo-AUTh &  1608.97 &  1409.95 &  1242.37 &  5596 &  5748 &  5672 \\
    \multirow{-6}{*}{ 2} & \cellcolor{Gray}F: cWB & \cellcolor{Gray}N/A & \cellcolor{Gray}N/A & \cellcolor{Gray}N/A & \cellcolor{Gray}N/A & \cellcolor{Gray}N/A & \cellcolor{Gray}N/A \\
    \hline
    \cellcolor{Gray} &  A: MFCNN &  885.46 &  472.96 &  355.83 &  37822 &  41251 &  39536 \\
    \cellcolor{Gray} & \cellcolor{Gray}B: PyCBC & \cellcolor{Gray}1734.43 & \cellcolor{Gray}1630.35 & \cellcolor{Gray}1577.18 & \cellcolor{Gray}149025$^{\ast\ast}$ & \cellcolor{Gray}146683$^{\ast\ast}$ & \cellcolor{Gray}147854$^{\ast\ast}$ \\
    \cellcolor{Gray} &  C: CNN-Coinc &  619.94 &  467.81 &  401.58 &  13628 &  14345 &  13986 \\
    \cellcolor{Gray} & \cellcolor{Gray}D: TPI FSU Jena & \cellcolor{Gray}1727.73 & \cellcolor{Gray}1577.77 & \cellcolor{Gray}1487.67 & \cellcolor{Gray}3862 & \cellcolor{Gray}3621 & \cellcolor{Gray}3742 \\
    \cellcolor{Gray} &  E: Virgo-AUTh &  1646.53 &  1494.98 &  1240.68 &  5450 &  5453 &  5451 \\
    \multirow{-6}{*}{\cellcolor{Gray}3} & \cellcolor{Gray}F: cWB & \cellcolor{Gray}1461.56 & \cellcolor{Gray}1359.78 & \cellcolor{Gray}1252.09 & \cellcolor{Gray}5247$^{\ast\ast\ast}$ & \cellcolor{Gray}N/A & \cellcolor{Gray}N/A \\
    \hline
      &  A: MFCNN &  1269.03 &  999.29 &  649.81 &  41942 &  46702 &  44322 \\
      & \cellcolor{Gray}B: PyCBC & \cellcolor{Gray}1722.43 & \cellcolor{Gray}1609.62 & \cellcolor{Gray}1544.33 & \cellcolor{Gray}162699$^{\ast\ast}$ & \cellcolor{Gray}163504$^{\ast\ast}$ & \cellcolor{Gray}163102$^{\ast\ast}$ \\
      &  C: CNN-Coinc &  960.70 &  620.85 &  0.00 &  12489 &  12431 &  12460 \\
      & \cellcolor{Gray}D: TPI FSU Jena & \cellcolor{Gray}257.87 & \cellcolor{Gray}0.00 & \cellcolor{Gray}0.00 & \cellcolor{Gray}3540 & \cellcolor{Gray}3487 & \cellcolor{Gray}3514 \\
      &  E: Virgo-AUTh &  1608.71 &  1400.30 &  1091.77 &  5462 &  5571 &  5516 \\
    \multirow{-6}{*}{ 4} & \cellcolor{Gray}F: cWB & \cellcolor{Gray}1406.88 & \cellcolor{Gray}1331.90 & \cellcolor{Gray}1229.14 & \cellcolor{Gray}4996$^{\ast\ast\ast}$ & \cellcolor{Gray}N/A & \cellcolor{Gray}N/A \\
    \hline
\end{tabularx}

    \caption{A summary of the analysis results for all submissions and all datasets. The columns labeled ``Sensitivity'' give the values for the sensitive distance at the three \acrshort{far}s $10^2$ per month, $10^1$ per month, and $10^0$ per month rounded to the second decimal place. The values lie on the lines in \autoref{fig:sens}. The columns labeled ``Runtime'' list the time for evaluation of the foreground and background set in seconds, respectively. The runtime column labeled ``average'' lists the mean time obtained from evaluating the foreground and background data. Entries labeled ``N/A'' are not available, because they were not measured. The \pycbc times labeled with $^\ast$ are only approximations. The analysis did not run on the challenge hardware but made use of a compute cluster. Shown times are the result of scaling the computational costs to 16 CPU cores. The \pycbc times labeled with $^{\ast\ast}$ are approximations obtained in the same manner as the approximations labeled with $^\ast$, but make use of a larger filter bank. The times of the \cwb group marked with $^{\ast\ast\ast}$ are approximations derived from dividing the CPU core-seconds reported by the search by $16$ to normalize it to the challenge hardware.}
    \label{tab:res}
\end{table*}

We find that the machine learning algorithms from the \jena group presented in \autoref{sec:submission-jena} and the \virgo group presented in \autoref{sec:submission-auth} are very close in sensitivity for datasets 1, 2, and 3. The submission from the \jena group reaches a slightly higher sensitive distance at all \acrshort{far}s for all of these three datasets. However, the \virgo submission retains $\geq 90\%$ of the sensitive distance achieved by the \jena submissions for \acrshort{far}s $\geq2$ per month. At lower \acrshort{far}s the gap widens but the individual sensitivities carry large uncertainties due to low number statistics. For higher \acrshort{far}s this gap narrows to a separation of roughly $4\%$ at a \acrshort{far} of $1000$ per month. We suspect that the difference between the two approaches is on the order that could be explained by different initializations of the training procedure.

On dataset 4 the submission from the \virgo group manages to maintain a stable sensitivity for the full range of tested \acrshort{far}s. The submission from \jena, on the other hand, is dominated by background triggers and seemingly struggles to adjust to the non-Gaussian noise characteristics. For high \acrshort{far}s the sensitivity is on a similar scale as the submission from the \virgo group and as was observed on previous datasets, backing up the hypothesis that rejecting background triggers is the main problem. This is surprising, as both algorithms were optimized on dataset 4 but performed similarly only on datasets 1 to 3. One reason for this result may be the neural network architectures used by the different groups. The \virgo group uses a very deep ResNet that may be better suited to represent non-Gaussian noise artifacts. The architecture from the \jena group is a more straightforward convolutional architecture that may be limited in its ability to learn appropriate parameters.

The algorithms from the \mfcnn group presented in \autoref{sec:submission-mfcnn} and the \cnn group presented in \autoref{sec:submission-cnn} also show similarities in sensitivity. Both are significantly less sensitive than the leading machine learning submission on all datasets. For datasets 1, 2, and 3, the \mfcnn contribution achieves $32.5\%$, $30.8\%$, and $23.5\%$ of the sensitive distances of the leading machine learning contribution, respectively. The \cnn submission reaches $42\%$, $25.5\%$, and $27\%$ of the sensitivity of the leading machine learning contribution at the point of farthest separation. For dataset 4 the submission from the \mfcnn and \cnn groups do comparatively better. They retain $\geq 68\%$ and $\geq 50\%$ of the sensitive distance of the leading machine learning submission down to a \acrshort{far} of $10$ per month, respectively. At a \acrshort{far} of $1$ per month the \cnn submission does not detect any signals, whereas the \mfcnn still retains $60\%$ of the sensitivity of the leading machine learning contribution.

On the first three datasets one can observe a steep gradient of the sensitivity curves at varying \acrshort{far}s for the \mfcnn and \cnn submissions. At even higher \acrshort{far}s the curves level off again and return to a similar slope observed at low \acrshort{far}s. The sudden increase leads to the \mfcnn submission being more sensitive than the modeled \pycbc search by up to $15\%$ on dataset 3 for \acrshort{far}s $> 200$ per month. This behavior is not present in any of the other submissions and we were not able to find a clear explanation. However, we observe that both algorithms have different trigger rates on the foreground and background set. If the background is estimated from the foreground data only, the sensitivity of both algorithms drops sharply. All other algorithms are robust to this change. We show these sensitivity curves in \autoref{fig:sens_fg} of the appendix.

For all datasets we compare the leading machine learning submission to the submission from \pycbc presented in \autoref{sec:submission-pycbc}. We also compare it to the submission from \cwb presented in \autoref{sec:submission-cwb} for datasets 3 and 4. These two are traditional, state-of-the-art search algorithms that have already been used successfully in past observation runs~\citep{LIGOScientific:2016fbo, LIGOScientific:2016aoc, LIGOScientific:2021djp}.

For dataset 1 we find that the machine learning search is able to achieve between $94\%$ and $99\%$ of the sensitivity obtained with PyCBC. These results are remarkably close and improve significantly on the findings from \citep{Schafer:2021cml}, which targeted a very similar dataset. However, the gap between the machine learning detection algorithm and the \pycbc search widens for lower \acrshort{far}s. Therefore, we expect that the \pycbc contribution will be able to attribute a substantially higher significance to many events. This is amplified by the ability of \pycbc to trivially increase the amount of data that can be used for background estimation by introducing time-slides between detectors~\citep{Usman:2015kfa, Schafer:2021cml}.

For dataset 2 the leading machine learning contribution gets even closer to the traditional algorithm from the \pycbc group. At low \acrshort{far}s $\leq 20$ per month it retains $\geq 93.5\%$ of the sensitivity achieved by the \pycbc submission. For high \acrshort{far}s $\geq 200$ per month it even manages to outperform the \pycbc submission and is up to $1.5\%$ more sensitive.

From dataset 2 to dataset 3 all submissions experience a slight increase of the measured sensitive distance. This may be surprising at first but can be explained by the distribution of the effective spin. For dataset 3 the spin orientations are distributed isotropically, which causes the average effective spin to be smaller than in dataset 2. This leads to few systems with large effective spin. The \pycbc search gains up to $3\%$ in sensitivity at low \acrshort{far}s, although it loses about $1\%$ in sensitivity at high \acrshort{far}s. A similar change can be observed in the submission from \jena. Since both the leading machine learning contribution and the \pycbc search gain similar amounts of sensitivity from dataset 2 to dataset 3 the comparison between the two does not change substantially. The submission from the \jena group is now up to $2.5\%$ more sensitive at high \acrshort{far}s and still about $6\%$ less sensitive at low \acrshort{far}s. The \virgo, the \mfcnn, and the \cnn submissions increase their sensitive distance by a larger fraction, suggesting that they benefit more from the signal population being closer to the distribution of signals in their training set. Dataset 3 is also the first dataset for which results from the \cwb search are available. We find that \cwb retains $\geq 80\%$ of the sensitive distance obtained by \pycbc over all tested \acrshort{far}s. Subsequently the leading machine learning submission achieves a sensitive distance greater by $15\%$ to $23\%$ over the range of tested \acrshort{far}s.

For dataset 4 the leading machine learning contribution now comes from the \virgo group. Compared to \pycbc their algorithm retains $\geq 87\%$ of the sensitivity down to a \acrshort{far} of $10$ per month. For smaller \acrshort{far}s the sensitivity gap widens quickly. At a \acrshort{far} of $1$ per month the machine learning search achieves $70\%$ of the sensitivity of \pycbc. The \cwb submission evolves similarly to \pycbc and retains $\geq 79\%$ of the sensitive distance. At high \acrshort{far}s the leading machine learning search manages a sensitive distance up to $27\%$ larger than that of \cwb. For low \acrshort{far}s the sensitive distance falls off quicker than that of \cwb. At a \acrshort{far} of $1$ per month the \cwb search is $12.5\%$ more sensitive than the \virgo submission. For lower \acrshort{far}s we expect this difference to become larger, as the production level search algorithms are tuned for lower \acrshort{far}s than tested in this work. In comparison to the sensitivity difference on dataset 3 the machine learning submission from \virgo does not retain as much sensitivity on real noise as the \pycbc or \cwb submissions.

The results on dataset 1 demonstrate that machine learning detection algorithms are already capable of rivaling traditional search algorithms for simulated data at \acrshort{far}s $\geq 1$ per month. A previous study~\citep{Schafer:2021cml} had identified the capability of machine learning searches to build an internal representation of the signal morphology as the main problem to achieve comparable sensitivities to traditional algorithms. Such a signal representation would allow the algorithms to compare detections in multiple detectors and require them to be consistent. The two leading machine learning algorithms in this challenge seem to have overcome this limitation, at least for high \acrshort{far} detections.

For dataset 2 we expected machine learning searches to decline in sensitivity more strongly than traditional searches. This expectation was provoked by the short duration of data that is processed by most machine learning searches at each step. As the signals injected into dataset 2 are of longer duration than those used in dataset 1, the machine learning algorithms inherently lose some amount of sensitivity due to considering only small parts of the signal. We estimate this loss to account for at most a $1\%$ difference in sensitivity. However, we observe the opposite effect for the two leading machine learning algorithms, which get even closer in sensitivity to the \pycbc submission compared to dataset 1. This may be caused by the distribution of signals in the training data used for the machine learning algorithms. Since both algorithms optimized for dataset 4, most signals in the training data will have non-zero spin. Therefore, the challenge set for dataset 2 is closer in nature to the training data, which may have introduced a bias that leads to higher sensitivities for spinning systems or in other words a slightly reduced sensitivity to non-spinning systems.

Dataset 3 was intended to test if machine learning searches are capable of outperforming traditional algorithms for precessing systems and signals carrying higher order mode information. We do not find substantial evidence in support of this hypothesis from the sensitivity curves. However, the challenge set 3 contains only very few signals with strong evidence for precession and higher order modes, as most signals are still relatively short. The impact on the overall sensitivity from these signals is, therefore, minor. Surprisingly, the leading machine learning search is still on par with \pycbc and manages to be significantly more sensitive even at the lowest tested \acrshort{far}s than the unmodeled \cwb search. It must be noted that the \cwb submission was not optimized for the parameter space used in this challenge. We, thus, expect this gap to narrow if more effort were to be used to tune the \cwb pipeline.

The change in the relative difference in sensitivity between the \pycbc submission and the leading machine learning contribution, as well as the change in difference to the \cwb submission, from dataset 3 to dataset 4 suggests that many machine learning algorithms currently used by the community are not yet capable of treating real noise as well as sophisticated traditional algorithms. We suspect that one major factor may be non-Gaussian noise artifacts that are misclassified as signals by machine learning algorithms, while the traditional searches excise them from the data or reject them on other bases. Another reason may be the non-stationary character of the noise that may lead to different sensitivities at different times. However, this would have also been a factor in dataset 3, where the \acrshort{psd}s used to simulate the noise change over the duration of the challenge set. However, since the leading machine learning search does retain sensitivity at all \acrshort{far}s it must have learned to reject most non-Gaussian noise artifacts, which is in line with expectations from studies carried out at higher \acrshort{far}s~\citep{George:2017pmj, Gebhard:2019ldz, Krastev:2020skk, Wei:2020sfz}.

\subsection{Found and missed injections}
We generate found-missed plots for all submissions and show a few selected ones. The ones not included in this paper can be found in the associated data release~\citep{github}. These plots highlight specific areas in parameter space where the machine learning searches are already competitive and those where more work is required. Specifically, we provide plots for chirp-mass $M_c$ versus decisive effective chirp-distance $D_{c,\text{eff}}$, $\tau_0$ versus mass-ratio $q$, and the effective precession spin $\chi_p$~\citep{Schmidt:2014iyl} versus inclination with respect to the line of sight $\theta_{jn}$. To first order $\tau_0$ is the time to merger from the lower frequency cutoff of the waveform~\citep{Cokelaer:2007kx, Maggiore:2007ulw}. The decisive effective chirp-distance is a measure for how strong the signal can be observed in the detector that has the worse sensitivity due to source location and orientation. The effective chirp-distance is the chirp-distance at which a source with the same intrinsic parameters and sky location but an optimal orientation would have been observed from at the same amplitude as the injected signal. The decisive effective chirp-distance is then the larger of the two effective chirp-distances from the two detectors. Therefore, the $M_c/D_{c,\text{eff}}$ plot informs about the ability to detect signals as a function of the \acrshort{snr} in the detector that is less sensitive to the signal. We also include information on the ranking statistic like quantity returned for each detected event, to highlight how strongly it is correlated to the \acrshort{snr}. The $\tau_0$ versus $q$ plot highlights how well long and short duration signals are recovered. It also gives information on the mass ratio, which is an important parameter for the strength of precession effects. The main plot used to determine the impact precession effects and higher order modes have on the detectability of signals is the $\chi_p$ versus $\theta_{jn}$ plot.

In \autoref{fig:ds1md} we show the found injections from dataset 1 in the $M_c$-$D_{c,\text{eff}}$-plane for the \pycbc and \jena submissions, respectively. Both plots clearly show that closer injections are generally attributed a higher confidence to be a real signal. This indicates that the ranking statistic like quantities for both algorithms are actually correlated with the signal strength. Similar correlations can be observed for all submissions. From \autoref{fig:ds1md} we find that the chirp-mass distribution from the \jena submission favors chirp-masses in the region $M_c\in[20, 35]$, which is not true for the \pycbc submission. We attribute this bias to the training set, which contained signals drawn from the distributions used for dataset 3 and 4. The probability distribution of the chirp-mass for these sets is shaped such that about $51\%$ of signals are being drawn from the mass range \SI{20}{M_\odot} to \SI{35}{M_\odot}. A similar bias is not so evident for the other machine learning submissions but may be masked by other effects. The \pycbc submission uses a uniform prior on the chirp mass and thus avoids this bias.

In \autoref{fig:ds2tq} we compare the found injections from dataset 2 in the $\tau_0$-$q$-plane for the \pycbc and \jena submissions. The plots show that the two searches are competitive in the comparable mass region and identify similar signals. The main difference between the two searches can be observed in the $\tau_0$ distribution of found signals. Most of the signals with large values for $\tau_0$, i.e. long duration signals, are missed by the \jena submission. These crucially include many signals that the \pycbc submission identifies with relatively high confidence. Therefore, the short duration of the input windows used by the \jena submission still seem to be a limiting factor for the sensitivity. This limitation will likely be more severe if longer duration signals from sources like \acrshort{bns} or \acrshort{nsbh} systems were considered.

In \autoref{fig:ds3chitheta} we compare the $\theta_{jn}$ and $\chi_p$ values of the injections from dataset 3 that are found by one algorithm but missed by the other. The two algorithms come from the \pycbc group and the \jena group. If either algorithm adapted better to signals with strong precession or higher order modes content, we would expect to see a clustering from that search in the scatter plot. However, we do not observe this clustering, which backs up our observation from the sensitivity curves that the machine learning algorithm from the \jena group has not learned a better representation of precessing systems or signals with higher order mode content than the modeled \pycbc search, which only includes non-precessing signals in its template bank. However, the amount of impact precession or higher order modes have on the detectability of short duration signals used in this study are small. A real test of this hypothesis would require the analysis of long duration signals.

\subsection{Runtimes}
All runtimes in this section are given in terms of wall-clock times obtained on equivalent hardware, which is listed in \autoref{tab:hardware}. The  runtimes are largely independent of the dataset for all submissions. We, therefore, discuss them only in summary. An overview of the times can be found in \autoref{tab:res}. They were measured by applying each algorithm to the foreground and background of each challenge set. We report the time between the algorithm call and it returning. To avoid bottlenecks, all files were transferred to the local storage of the individual compute nodes before calling the algorithm. The output was also written to said local storage and transferred back only after the algorithm returned. It should be noted that the runtimes are heavily dependent on the amount of optimization of the algorithms. The main objective for this challenge was the sensitivity and not the runtime.

The \pycbc and \cwb submissions are exceptions as their runtimes were not measured on the same hardware. Instead they were run on compute clusters making heavy use of parallelized work over multiple CPUs. The times reported here are approximations by normalizing the compute time to 16 CPU cores available in the compute nodes used for this challenge. Furthermore, for the evaluation of dataset 1 \pycbc used a different template bank than those for dataset 2 to 4 was used. This bank was substantially smaller, resulting in faster evaluation. \cwb times were reported to us only on the foreground data in CPU core seconds.

We find that of the machine learning algorithms the submission from the \jena group is the fastest, evaluating an entire month of archival data in about \SI{1}{\hour}. It utilizes a single GPU when evaluating the network. The second fastest algorithm is the submission from the \virgo group. It evaluates a month of data in \SI{1.5}{\hour} on a single GPU and is thus about $50\%$ slower than the fastest algorithm. Notably, the two fastest algorithms are also the two most sensitive machine learning searches. The algorithm from the \cnn group requires almost \SI{4}{\hour} on a single GPU to evaluate the same amount of data but is significantly less sensitive. However, none of these algorithms are limited by the GPU performance. The differences in execution time can be mainly attributed to the difference in optimization of the pre-processing steps. The submission from the \mfcnn group on the other hand does not apply any pre-processing directly. They instead use a neural network to carry out this computation. They operate on all 8 available GPUs and manage to evaluate the month of data in \SI[parse-numbers=false]{\approx 11.5}{\hour}.

For dataset 1 the \pycbc submission has a runtime comparable to that of the submission from the \virgo group. On all other datasets it requires roughly \SI{43}{\hour} to evaluate the month of data. The large difference in runtime between the datasets is caused by the smaller template bank that is used only for dataset 1. Contrary to the machine learning algorithms, the \pycbc submission did not utilize GPUs and ran on CPUs only. However, \pycbc is a production level search pipeline and as such has been optimized to run on CPUs. It is not limited by the pre-processing but rather by the matched filter operation. It should be noted that \pycbc is still the most sensitive search presented here and gains in computational efficiency could be obtained by reducing the number of templates. This would effectively trade off search sensitivity for lower computational cost.

The \pycbc submission is implemented on the CPU as a GPU implementation is inherently more difficult to optimize. GPUs, on the other hand, are usually far more efficient from a cost to performance and energy to performance standpoint~\citep{Dhurkunde:2021csz}. One advantage of machine learning algorithms is that they make use of well optimized libraries such as PyTorch~\citep{pytorch_software} or TensorFlow~\citep{tensorflow_software} that utilize GPUs for their computations. This makes the implementation of search algorithms on GPUs relatively straightforward and allows researchers to focus on optimizing the sensitivity of their algorithm rather than having to spend time on optimizing the algorithmic implementation. 

The runtimes in this challenge are measured under the assumption that the lowest required \acrshort{far} is $1$ per month. In a real search lower \acrshort{far}s are beneficial especially for rare signals. Therefore, most traditional searches are tuned to be most sensitive at \acrshort{far}s well below the level tested in this challenge. \pycbc for instance can extend its background by introducing time-slides~\citep{Usman:2015kfa}, thereby potentially lowering the \acrshort{far}s of detected events. This process is a trivial operation that requires a fraction of the computational cost of the actual filtering stage. If machine learning algorithms are not specifically designed to allow for a similar approach, lowering the \acrshort{far}s of detections requires multiple complete re-evaluations of the time-shifted data. This would in turn lead to a linear increase in the computational cost, i.e. lowering the potential \acrshort{far} of an event by an order of magnitude would lead to an order of magnitude increase in the computational cost.

\section{Conclusions}\label{sec:conclusions}
In this paper we have presented the results of the first Machine Learning Gravitational Wave Search Mock Data Challenge (MLGWSC-1). The study compiled curves showing the sensitive distances from $4$ different machine learning submissions and compared them to $2$ state-of-the-art traditional search algorithms; the modeled PyCBC~\citep{Usman:2015kfa} pipeline and the unmodeled coherent wave burst search~\citep{Klimenko:2015ypf,CWBwavelet:2004km}. We established a common dataset and means for evaluation. We hope that other researchers will continue to make use of the resources presented in this work to allow for quantitative comparisons between different machine learning approaches and to traditional filtering techniques. As research continues and machine learning search algorithms become more sensitive, we want to motivate other groups to host new challenges, focusing on other parts of parameter space or targeting different observing strategies.

The key observations of this challenge are:
\begin{enumerate}
    \item Machine learning search algorithms are competitive in sensitivity compared to state-of-the-art searches on simulated data and the limited parameter space explored in this challenge.
    \item Most of the tested machine learning algorithms struggle to effectively handle real noise, which is contaminated with non-Gaussian noise artifacts.
    \item Traditional search algorithms are capable of detecting signals at lower \acrshort{far}s, thus making detections more confident.
    \item The tested machine learning searches struggle to identify long duration signals.
\end{enumerate}
Therefore, the main challenges for current machine learning searches are the operation on real noise, the confidence in detections due to comparatively high \acrshort{far}s, and the detection of long duration signals. The last of those three is a major hurdle to confidently detect signals from \acrshort{bns} and \acrshort{nsbh} systems. Improvements in any of these fields would be beneficial. Specifically, we identify the following key research areas:
\begin{enumerate}
    \item Improve the ability to compare signal parameters, or representations thereof, between detectors to check for consistency and reject noise artifacts.
    \item Improve the ability to calculate large amounts of background, for instance by designing algorithms that can trivially evaluate time-slides of the input data.
    \item Increase the duration of data that is processed by machine learning algorithms to enable the detection of long duration signals.
\end{enumerate}

This challenge shows the potential of machine learning algorithms to act as \acrshort{gw} detection pipelines. We have shown that these algorithms are competitive in a realistic scenario to state-of-the-art searches today. They operate at low computational cost and allow for a trivial implementation of the algorithms on highly efficient GPUs, rather than relying on CPUs. We believe that this work justifies more research on this topic, especially in areas where machine learning may have a tangible impact on the rapid identification of \acrshort{gw}s.

However, we do acknowledge that the research carried out here operates on a limited parameter space. Moreover, the targeted parameter space is not the computationally expensive part of the search space of traditional searches. About $1\%$ of the total size of the template bank used in \citep{Nitz:2021zwj} is dedicated to the area this study searches. To have the greatest impact on real searches machine learning algorithms need to be extended to target either the low mass region, where signals are long and the computational cost of matched filtering rises rapidly, or the high mass region where signals and noise artifacts are difficult to distinguish.

We also want to mention that we did not receive a submission utilizing one of the most promising neural network architectures for \acrshort{gw} detection of the recent past. A WaveNet based architecture, that uses dilated convolutions, has been reported to do well for this kind of task~\citep{Gebhard:2019ldz, Schmitt:2019aaa, Wei:2020ztw}. We also did not receive submission based on many other neural network architectures that have been used in the past, such as autoencoders~\citep{Shen:2019ohi, Morawski:2021kxv, Moreno:2021fvp, Bacon:2022lsm}, inception networks~\citep{Dreissigacker:2020xfr, Schafer:2020kor}, or two dimensional convolutions that analyze time-frequency decompositions~\citep{Wei:2020sfz}. We hope that some of these approaches will be adopted to the requirements of this challenge and evaluated on the datasets presented here, to allow for a quantitative comparison.

Future mock data challenges could target longer duration signals, concentrating on \acrshort{bns} and \acrshort{nsbh} systems. These are potentially EM-bright and would, therefore, be of particular interest. Furthermore, these signals stem from regions of parameter space where traditional searches are computationally expensive to run. For even longer signals, sub-solar mass black holes could be targeted. Existing modeled searches in those regions make use of several million templates and are computationally limited ~\citep{Nitz:2022ltl}. Another avenue may be very massive systems, which can be difficult to distinguish from noise artifacts. Finally, we recommend that future mock data challenges drop the notion of a foreground and background set and only provide data files containing injections. This would eliminate further sources of error and be more true to a realistic application, where no true \acrshort{gw}-free background exists.

\section{Acknowledgements}
We want to thank Narenraju Nagarajan and Pascal Müller for their valuable scientific input and contributions to the code of this challenge.

We acknowledge the Max Planck Gesellschaft and the Atlas cluster computing team at Albert-Einstein Institut (AEI) Hannover for support.

O.Z. thanks the Carl Zeiss Foundation for the financial support within the scope of the program line "Breakthroughs".

The MFCNN team members would like to acknowledge that the submission was supported by the Peng Cheng Laboratory Cloud Brain (No. PCL2021A13).

The \virgo team members would like to acknowledge the support provided by the IT Center of the Aristotle University of Thessaloniki (AUTh) throughout the progress of this work, as results presented in this work have been produced, in part, using the AUTh High Performance Computing Infrastructure and Resources, and thank the COST network CA17137 ``G2Net'' for support. P.I. acknowledges support by the European Research Council (ERC) under the European Union's Horizon 2020 research and innovation programme under grant agreement No. 759253.

The cWB team gratefully acknowledges the computational resources provided by LIGO-Virgo. This material is based upon work supported by NSF’s LIGO Laboratory, which is a major facility fully funded by the National Science Foundation. This research has made use of data, software and/or web tools obtained from the Gravitational Wave Open Science Center, a service of LIGO Laboratory, the LIGO Scientific Collaboration and the Virgo Collaboration. The work by S. K. was supported by NSF Grants No. PHY 1806165 and PHY 2110060.

This publication is based upon work from COST Action CA17137, supported by COST (European Cooperation in Science and Technology).

EAH is supported by Laboratory Directed Research and Development (LDRD) funding from Argonne National Laboratory, provided by the Director, Office of Science, of the U.S. Department of Energy under Contract No. DE-AC02-06CH11357, and the U.S. National Science Foundation Awards OAC-2209892 and OAC-1931561.

CM is supported by the Science and Technology Research Council (grant no. ST/V005634/1) and the European Cooperation in Science and Technology (COST) action CA17137.

F.O. was supported by the Max Planck Society's Independent Research Group Programme.

This research has made use of data or software obtained from the Gravitational Wave Open Science Center (gw-openscience.org), a service of LIGO Laboratory, the LIGO Scientific Collaboration, the Virgo Collaboration, and KAGRA. LIGO Laboratory and Advanced LIGO are funded by the United States National Science Foundation (NSF) as well as the Science and Technology Facilities Council (STFC) of the United Kingdom, the Max-Planck-Society (MPS), and the State of Niedersachsen/Germany for support of the construction of Advanced LIGO and construction and operation of the GEO600 detector. Additional support for Advanced LIGO was provided by the Australian Research Council. Virgo is funded, through the European Gravitational Observatory (EGO), by the French Centre National de Recherche Scientifique (CNRS), the Italian Istituto Nazionale di Fisica Nucleare (INFN) and the Dutch Nikhef, with contributions by institutions from Belgium, Germany, Greece, Hungary, Ireland, Japan, Monaco, Poland, Portugal, Spain. The construction and operation of KAGRA are funded by Ministry of Education, Culture, Sports, Science and Technology (MEXT), and Japan Society for the Promotion of Science (JSPS), National Research Foundation (NRF) and Ministry of Science and ICT (MSIT) in Korea, Academia Sinica (AS) and the Ministry of Science and Technology (MoST) in Taiwan.

\bibliography{bibliography}

\appendix
\section{Sensitivity plot foreground}
\begin{figure*}
    \centering
    \includegraphics[width=0.98\textwidth]{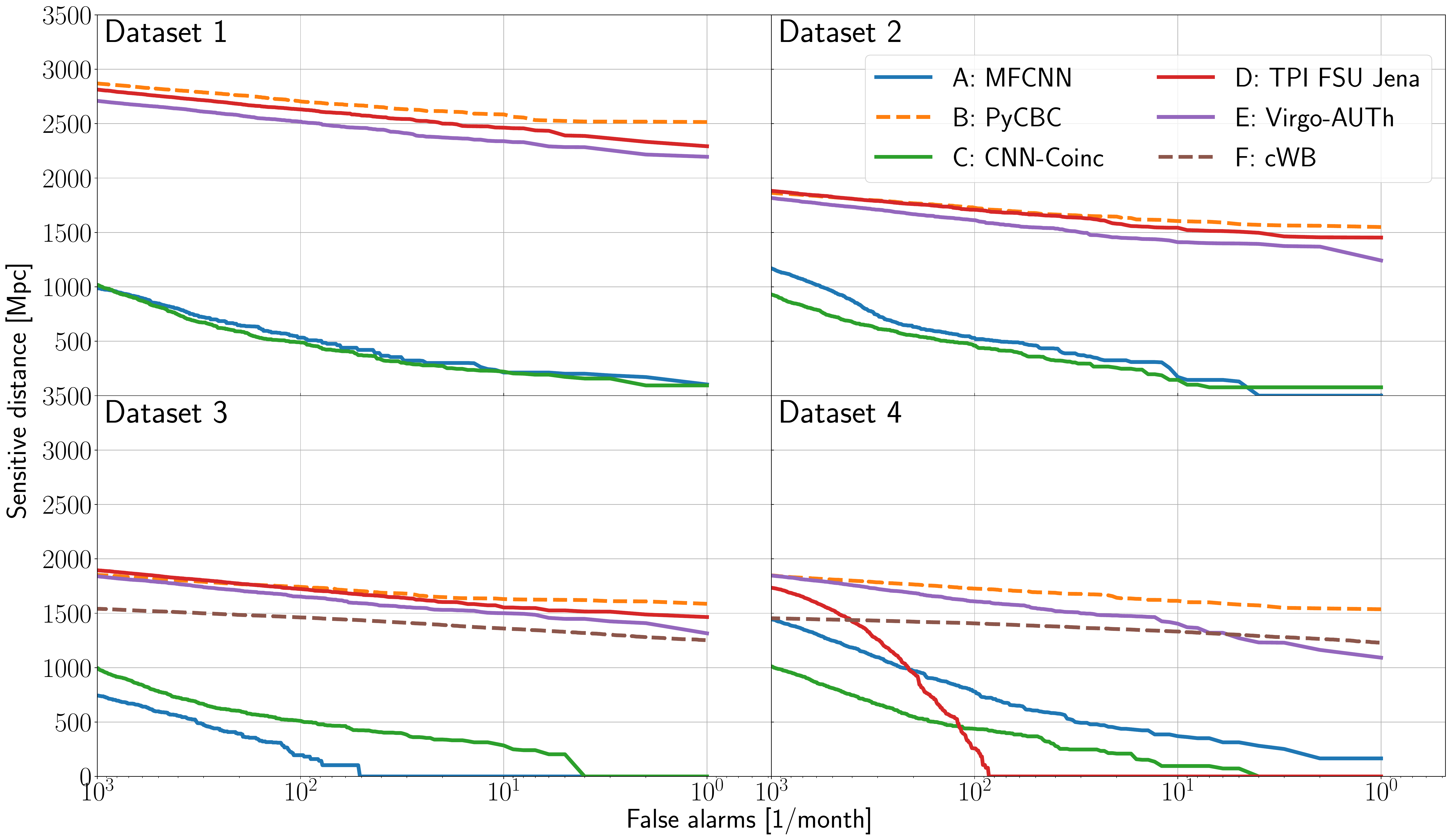}
    \caption{The sensitive distances of all submissions and all four datasets as functions of the \acrshort{far}. The sensitive distances are calculated using only the data from the foreground file. The \acrshort{far} is determined from the false positives on that data. Submissions that made use of a machine learning algorithm at their core are shown with solid lines, others with dashed lines. This figure differs from \autoref{fig:sens} as the algorithms from \mfcnn and \cnn behave differently on the foreground and the background.}
    \label{fig:sens_fg}
\end{figure*}

\section{Found-Missed plots}
\begin{figure*}
    \centering
    \includegraphics[width=0.9\textwidth]{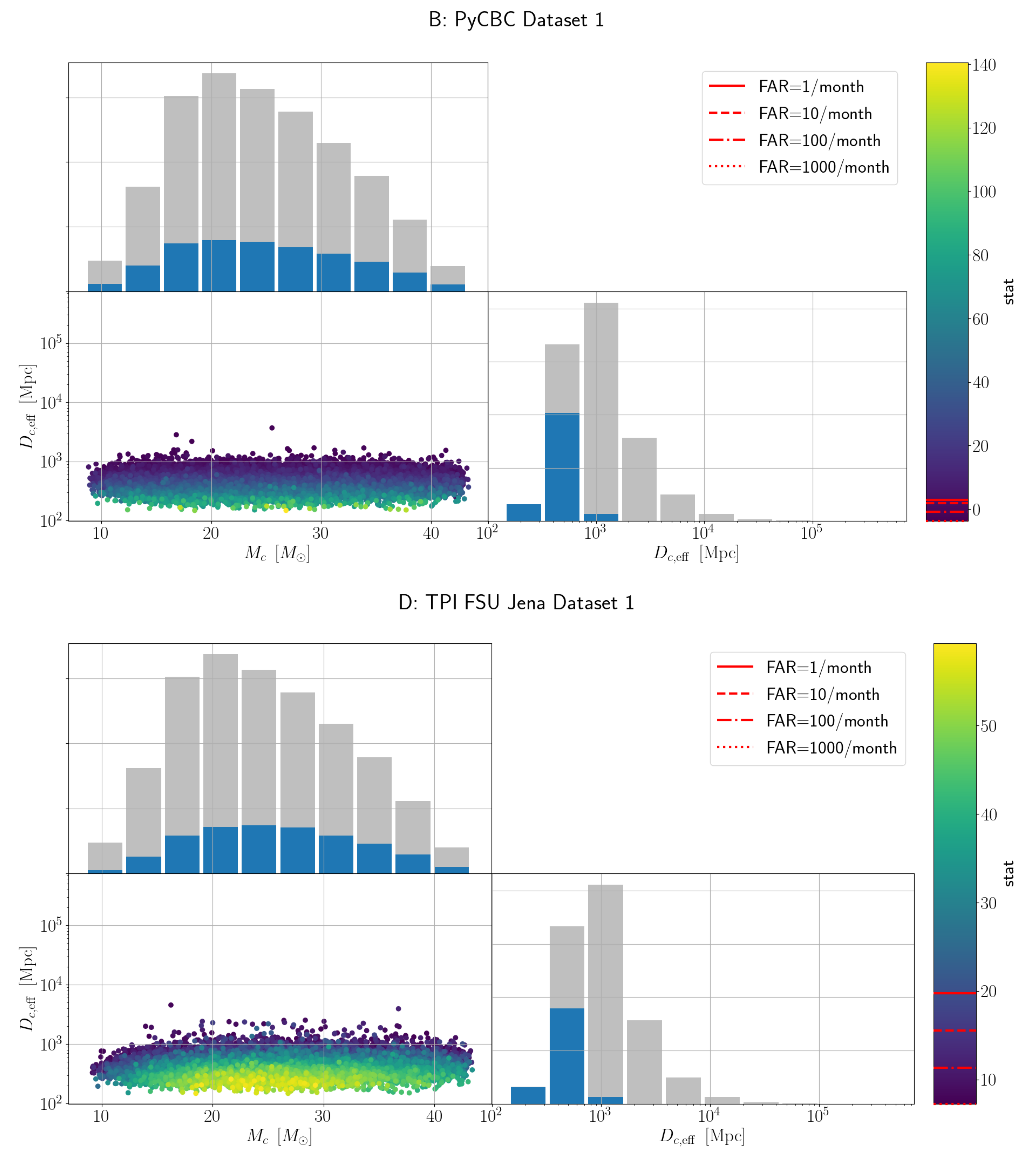}
    \caption{The injections from dataset 1 identified by the \pycbc and \jena submissions with a \acrshort{far} $\leq 10^3$ per month in the chirp-mass $M_c$ versus decisive effective chirp-distance $D_{c,\text{eff}}$ plane. The blue bars in the histograms show the one dimensional marginal distributions of the found injections. The gray bars show the distribution of injected signals, including those missed by the search. The color shows the ``stat'' value attributed to the injection by the algorithm. The red lines in the colorbar highlight the thresholds on the ``stat'' to achieve different \acrshort{far}s.}
    \label{fig:ds1md}
\end{figure*}

\begin{figure*}
    \centering
    \includegraphics[width=0.9\textwidth]{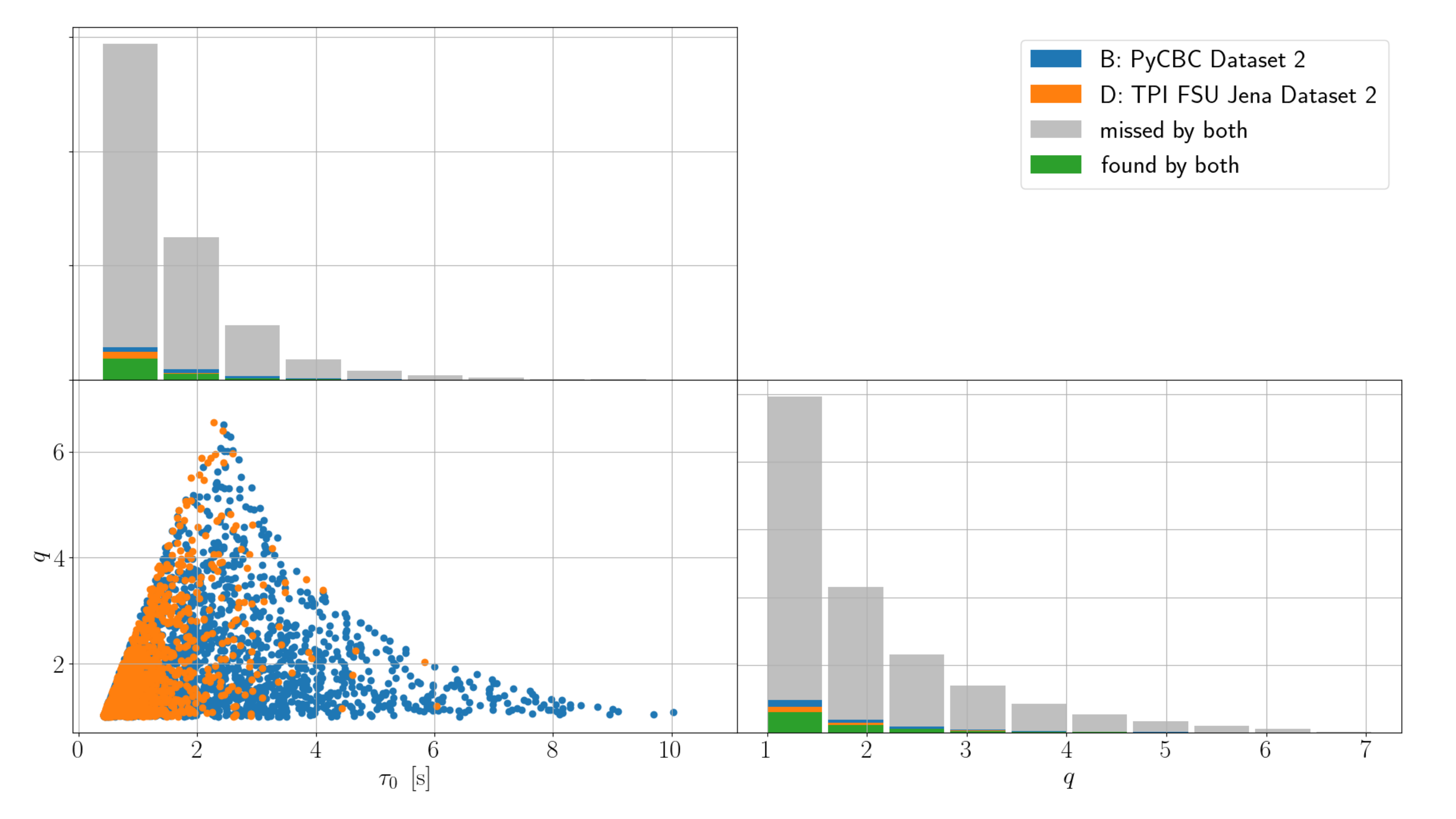}
    \caption{The injections from dataset 2 identified by the \pycbc and \jena submissions with a \acrshort{far} $\leq 10^3$ per month in the signal duration $\tau_0$ versus mass ratio $q$ plane. The scatter plot shows injections that are found only by one of the two algorithms. Injections that are missed or found by both are only shown in the 1D marginal distributions.}
    \label{fig:ds2tq}
\end{figure*}
\begin{figure*}
    \centering
    \includegraphics[width=0.9\textwidth]{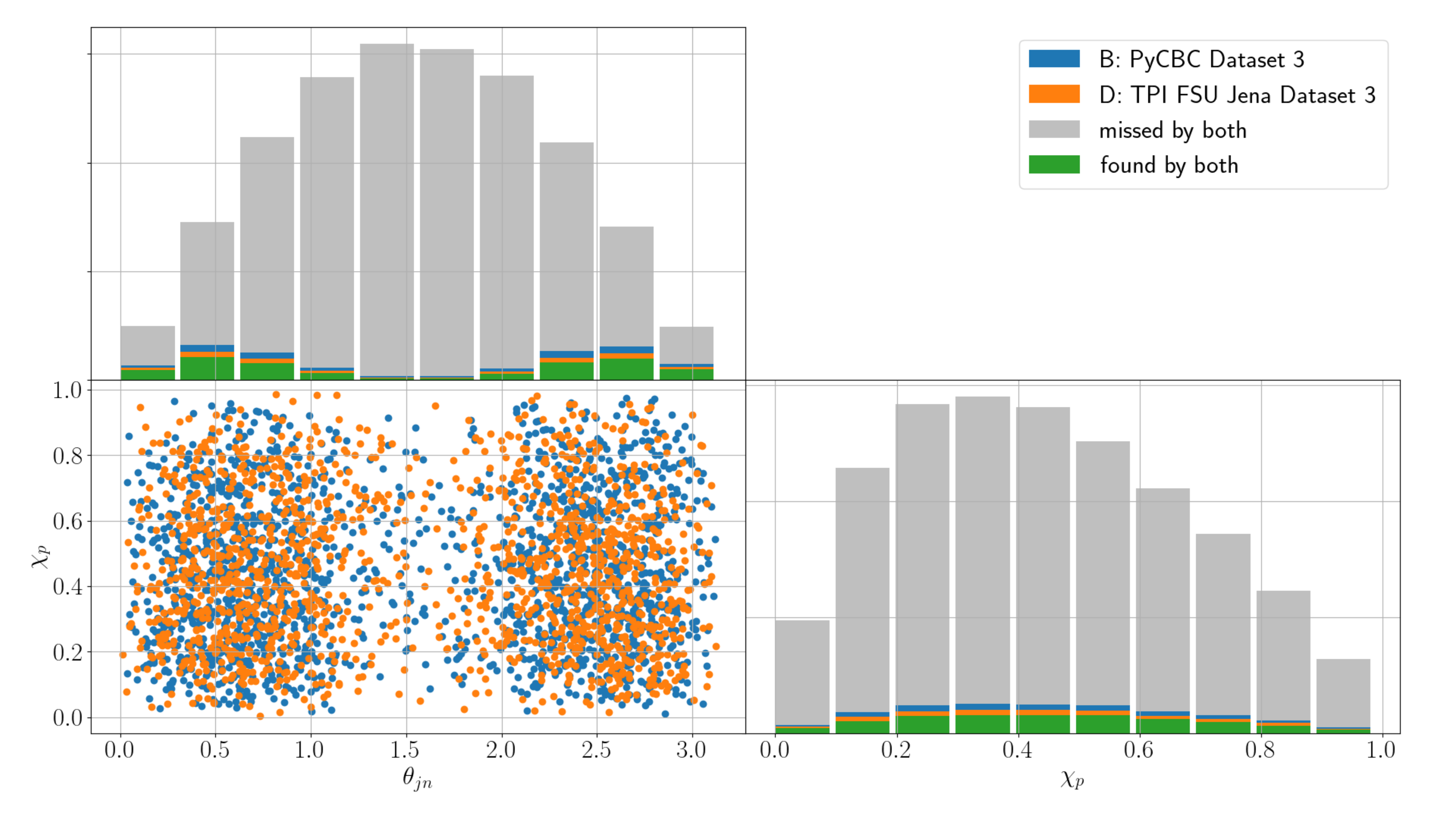}
    \caption{The injections from dataset 3 identified by the \pycbc and \jena submissions with a \acrshort{far} $\leq 10^3$ per month in the inclination to spin axis $\theta_{jn}$ to $\chi_p$ plane. The scatter plot shows injections that are found only by one of the two algorithms. Injections that are missed or found by both are only shown in the 1D marginal distributions.}
    \label{fig:ds3chitheta}
\end{figure*}
\end{document}